\documentclass[a4paper,preprintnumbers,floatfix,twocolumn,aps,prb,unsortedaddress,superscriptaddress,longbibliography]{revtex4-2}

\usepackage[para]{threeparttable} % for table footnotes
\usepackage{amsmath}
\usepackage{graphicx} 
\usepackage{booktabs}
\usepackage{caption}
\usepackage{subcaption}
\usepackage{nicefrac}
\usepackage{url}
\usepackage{hyperref}
\usepackage{miller}
\usepackage{bm}
\usepackage{enumitem}
\usepackage{color}
\usepackage{arydshln}
\usepackage{tabu}% http://ctan.org/pkg/tabu
\usepackage{xcolor}% http://ctan.org/pkg/xcolor
\usepackage{tabularx}% http://ctan.org/pkg/tabularx

\newcommand{\etal}{{\it et al. }}
\usepackage[paperwidth=210mm,paperheight=297mm,centering,hmargin=1.7cm,vmargin=2.5cm]{geometry}

\renewcommand{\selectlanguage}[1]{} % fixing errors in bibtex
\DeclareUnicodeCharacter{2264}{\ensuremath{\leq}}
\makeatletter
\newcommand{\dashrule}[1][black]{%
  \color{#1}\rule[\dimexpr.5ex-.2pt]{4pt}{.4pt}\xleaders\hbox{\rule{4pt}{0pt}\rule[\dimexpr.5ex-.2pt]{4pt}{.4pt}}\hfill\kern0pt%
}
\newcommand{\rulecolor}[1]{%
  \def\CT@arc@{\color{#1}}%
}
\makeatother

\begin{document}

\title{Radiation damage and phase stability of Al$_x$CrCuFeNi$_y$ alloys using a machine-learned interatomic potential}

\author{A. Fellman}
\thanks{Corresponding author}
\email{aslak.fellman@helsinki.fi}
\affiliation{Department of Physics, P.O. Box 43, FI-00014 University of Helsinki, Finland}
\author{J. Byggmästar}
\affiliation{Department of Physics, P.O. Box 43, FI-00014 University of Helsinki, Finland}
\author{F. Granberg}
\affiliation{Department of Physics, P.O. Box 43, FI-00014 University of Helsinki, Finland}
\author{F. Djurabekova}
\affiliation{Department of Physics, P.O. Box 43, FI-00014 University of Helsinki, Finland}
%\affiliation{Helsinki Institute of Physics, Helsinki, Finland}
\author{K. Nordlund}
\affiliation{Department of Physics, P.O. Box 43, FI-00014 University of Helsinki, Finland}

\date{\today}

\begin{abstract}
We develop a machine-learned interatomic potential for AlCrCuFeNi high-entropy alloys (HEA) using a diverse set of structures from density functional theory calculated including magnetic effects. The potential is based on the computationally efficient tabulated version of the Gaussian approximation potential method (tabGAP) and is a general-purpose model for molecular dynamics simulation of the HEA system, with additional emphasis on radiation damage effects. We use the potential to study key properties of AlCrCuFeNi HEAs at different compositions, focusing on the FCC/BCC phase stability. Monte Carlo swapping simulations are performed to understand the stability and segregation of the HEA and reveal clear FeCr and Cu segregation. Close to equiatomic composition, a transition from FCC to BCC is detected, following the valence electron concentration stability rule. Furthermore, we perform overlapping cascade simulations to investigate radiation damage production and tolerance. Different alloy compositions show significant differences in defect concentrations, and all alloy compositions show enrichment of some elements in or around defects. We find that, generally, a lower Al content corresponds to lower defect concentrations during irradiation. Furthermore, clear short-range ordering is observed as a consequence of continued irradiation.
\end{abstract}

\maketitle

\section{Introduction}
\label{sec:intro}

High-entropy alloys (HEA) are a class of materials that have traditionally been defined as consisting of five or more constituent elements in equal or near equal concentrations in a single solid solution~\cite{yeh2004nanostructured,cantor2004microstructural}. However, more recently it has become apparent that complex multi-component systems can exhibit phase separation or have tendencies to form ordered phases and the requirement for equimolar concentrations has been found restrictive~\cite{manzoni2013phase, miracle2017critical}. HEAs have shown promising properties such as very high stability at high temperatures, good corrosion resistance as well as advanced mechanical properties~\cite{yeh2004nanostructured, qiu2013microstructure,ng2014phase,luo2020selective,xue2022development}. A good portion of the face-centered cubic (FCC) HEAs studied to date include large fractions of cobalt. However, there are several aspects, such as price, ethical concerns of its mining, and high activation in radiation applications that make cobalt an unfavorable alloying element from a sustainability point of view. Therefore, there is considerable interest in finding Co-free HEA compositions with good properties. 

Al$_x$CrCuFeNi$_y$ alloys have been found to be a good candidate for a Co-free HEA system. Depending on the composition and manufacturing processes, the alloy can exhibit both FCC and body-centered cubic (BCC) phases as well as ordered phases~\cite{li2009thermodynamic,guo2013anomalous,ng2014phase,borkar2016combinatorial,shivam2023microstructural}. In Al$_x$CrCuFeNi$_2$ it was revealed that an increase in Al content facilitated the change from FCC to FCC/BCC coexistence~\cite{borkar2016combinatorial,su2020microstructure}. Conversely, in AlCrCuFeNi$_x$ alloys it was noticed that an increase in Ni content seemed to stabilize the FCC phase and in multiphase systems it increased the fraction of FCC compared to BCC~\cite{jinhong2012microstructure,luo2020selective,luo2021microstructural}. Additionally, there seems to be some sensitivity to Al content in regards to mechanical properties, such as hardness~\cite{guo2013anomalous,borkar2016combinatorial}. 

With good general interatomic potentials, molecular dynamics (MD) simulations can be used to efficiently study different compositions and their properties, providing valuable information for selecting candidate materials and accelerating property optimization. There have been some studies of AlCrCuFeNi alloys using MD~\cite{li2016atomic, li2016mechanical,wang2017investigation,zeng2019thermal,niu2021molecular,doan2021effects,nguyen2023plastic,nguyen2023cyclic,nguyen2024machining,doan2025mechanical}. These mostly focused on mechanical properties such as deformation mechanisms due to nanoindentation or tensile loading simulations. However, in these studies the interatomic potential used is a mix of different existing potentials with simple Morse pair potentials for the missing interactions, rendering it hard to validate these complex systems and findings. Hence, the unavailability of a reliable interatomic potential for AlCrCuFeNi has significantly limited the computational studies of the material. 

In recent years, machine learning (ML) potentials have become commonplace in MD simulations and their availability has increased rapidly. ML potentials have shown excellent accuracy compared to traditional analytical potentials~\cite{bartok_machine_2018,deringer_machine_2017}. Therefore, ML potentials can bridge the gap between \textit{ab initio} methods, such as density functional theory (DFT), and classical analytical models. However, many ML potentials frameworks with high-dimensional descriptors for the local atomic environments struggle with the addition of many different chemical species~\cite{darby_compressing_2022}, either in terms of accuracy, computational speed, or both. The issue can be mitigated by methods for dimensionality reduction or chemical embeddings~\cite{darby_compressing_2022,chen_universal_2022,fan_improving_2022}, although most many-element potentials are still too computationally expensive for large-scale simulations. Recently, it was proposed that instead of using complex descriptors, simple low-dimensional descriptors such as two-body, three-body, and scalar density descriptors can make the training of HEA potentials data efficient and accurate. Moreover, tabulation of the final model gives orders-of-magnitude speed up benefits~\cite{byggmastar_simple_2022}. 

An additional complication for interatomic potentials of Fe- and Ni-containing alloys is the need to account for magnetism. Since Fe and Ni are strongly ferromagnetic, alloys thereof also contain significant magnetic moments~\cite{schneeweiss_magnetic_2017}. Explicitly including magnetic degrees of freedom in the potential formalism is still challenging~\cite{rinaldi_noncollinear_2024}, especially for alloys~\cite{shenoy_collinearspin_2024}.Implicitly accounting for magnetism by fitting to spin-polarized DFT data is common practice for pure elements like Fe~\cite{mendelev_development_2003,dragoni_achieving_2018,byggmastar2022multiscale}, but for alloys it is still rarely attempted~\cite{lopanitsyna_modeling_2023,song_generalpurpose_2024}. In recent years MLIPs that include explicit treatment of magnetism have started to appear, mainly for iron, fixed-composition iron-based alloys and iron oxides~\cite{nikolov2021data,chapman2022machine,novikov2022magnetic,rinaldi_noncollinear_2024,shenoy_collinearspin_2024,bienvenu2025development}. To our knowledge, no attempt has been made to (implicitly or explicitly) consider magnetism in ML potentials for FCC HEAs. Neglecting magnetism completely for Al$_x$CrCuFeNi$_y$ alloys prohibits studies of phase stability and segregation, since without spin-polarized DFT the ground state of Fe is FCC instead of BCC. In EAM and MEAM alloy potentials~\cite{zhou_misfit-energy-increasing_2004,choi_understanding_2018}, the elemental potentials are fitted to experimental data and hence do reproduce the correct ground state without explicitly considering magnetism, but the accuracy and reliability for studying compositional phase transitions is unclear.

There has been significant interest in MD studies of radiation damage in HEAs. This is due to favorable properties in regards to radiation resistance, like lower defect concentrations, inhibited cluster growth and low swelling~\cite{granberg2016mechanism,wei2024revealing}. In AlCoCrFeNi alloys there have been studies of radiation damage both experimentally~\cite{xia2015irradiation,yang2016precipitation,yang2020structural,an2025enhanced} and computationally~\cite{wang2023molecular,ma2024effect}. However, as far as the authors are aware, in AlCrCuFeNi alloys neither experimental nor computational studies have been performed. Due to the high activation of cobalt, there is interest in finding Co-free alloy compositions for the use in radiation applications. Furthermore, most studies focus on single-phase materials, meaning the radiation response of dual-phase materials is less understood. 

In this work, we enable large-scale simulations of AlCrCuFeNi HEAs by developing and validating a general-purpose and computationally efficient ML potential, trained to magnetic DFT data. We then study the phase stability, grain boundary segregation, radiation damage, and short-range order of Al$_x$CrCuFeNi$_y$ alloys.

\section{Methods}
\label{sec:methods}

\subsection{Tabulated Gaussian approximation potential}

The potential developed in this work is made using the Gaussian approximation potential (GAP) method~\cite{bartok_gaussian_2010}. GAP is a machine-learning framework based on sparse Gaussian process regression and some combination of descriptors for encoding local atomic environments. More specifically, the potential developed here is a low-dimensional tabulated version (tabGAP) of GAP~\cite{byggmastar_simple_2022}. tabGAP allows for orders of magnitude speed up compared to the GAP potential, allowing for simulations of large system sizes. Furthermore, tabGAP has been used successfully to create potentials for BCC refractory HEAs~\cite{byggmastar_simple_2022,byggmastar_modeling_2021}, where it was found that the simple formalism of tabGAP has benefits when the amount of chemical species increases. 

The total energy for the tabGAP is described using the following equation:

\begin{equation}
\begin{split}
E_\mathrm{tot.} = E_{\mathrm{rep.}}  &+  \sum_{i<j}^N \delta^2_\mathrm{2b} \sum_s^{M_\mathrm{2b}} \alpha_{s} K_\mathrm{se} (r_{ij}, r_s)  \\ 
& + \sum_{i, j < k}^N \delta^2_\mathrm{3b} \sum_s^{M_\mathrm{3b}} \alpha_{s} K_\mathrm{se} (\bm{q}_{ijk}, \bm{q}_s) \\ 
& + \sum_{i}^N \delta^2_\mathrm{eam} \sum_s^{M_\mathrm{eam}} \alpha_{s} K_\mathrm{se} (\rho_i, \rho_s) ,
\end{split}
\label{eq:tabgap}    
\end{equation}
where the sum over $N$ refers to the sum over atoms and $M$ refers to selected sparsified subset of descriptor environments from training structures. All terms consist of pre-factors $\delta^2$, regression coefficients $\alpha_{s}$, and squared exponential kernels $K_\mathrm{se}$ which measure the similarity between known and unknown descriptor values. The two-body descriptor is simply the distance between two atoms. The three-body descriptor is a permutation-invariant vector~\cite{bartok_gaussian_2015} and the embedded atom method (EAM) density descriptor is a scalar pairwise summed radial function, as in standard EAM potentials~\cite{byggmastar_simple_2022,daw_embedded-atom_1984,finnis_simple_1984}.

Additionally, the model includes an external purely repulsive potential. The repulsive part is included in order to more accurately describe short-range interactions, needed for high-energy simulations. The repulsive term is a Ziegler-Biersack-Littmark-type (ZBL) repulsive potential:
\begin{equation}
E_{\mathrm{rep.}} = \sum_{i<j}^{N} \frac{1}{4 \pi \epsilon_0} \frac{Z_i Z_j e^2}{r_{ij}} \phi (r_{ij}/a) f_{\mathrm{cut}}(r_{ij}),
\end{equation}
where
\begin{equation}
    a = \frac{0.46848}{Z_i^{0.23} + Z_j^{0.23}}.
\end{equation}

The screening function $\phi (r_{ij}/a)$ was refitted to repulsive dimer data from all-electron DFT calculations~\cite{Nor96c} for each element pair. Additionally, the screened Coulomb potential is multiplied by a cutoff function to force it to zero well below the nearest-neighbour distance of the material, to avoid interfering with the near-equilibrium interactions described by the machine-learned part. The range of the cutoff was chosen to be $1.0$--$2.2$ Å to ensure smooth transition between the all-electron calculations and short-range DFT data that was part of the training data. 

After the initial training of the potential, the energy contributions of the different terms in Eq.~\ref{eq:tabgap} are tabulated onto low-dimensional grids and evaluated using cubic-spline interpolation. The repulsive and two-body terms are tabulated into a one-dimensional grid for each element pair, the three-body term into a three-dimensional grid for each unique element triplet, and the EAM term into two one-dimensional grids. Further details on the tabulation process can be found in Refs.~\citenum{byggmastar2022multiscale,byggmastar_simple_2022}. The simple descriptors of the tabGAP model allows for this tabulation, which gives the model a two-orders-of-magnitude increase in computational efficiency compared to GAP, making the potential more suitable for large-scale simulations. 

The GAP fitting includes several hyperparamaters used for the different descriptors of the potential. The cutoff radii of the descriptors were 5.2 Å for the two-body and the EAM-like scalar density descriptor, while 4.7 Å was chosen for the three-body descriptor. The number of sparse descriptor environments from the training structures $M$ for the two-body, EAM and three-body descriptors, were 20, 30 and 500 respectively. The sparse sampling method for all descriptors in this work used uniform sampling as implemented in the GAP framework. Another set of hyperparameters are the regularisation terms $\sigma$ for the energies, forces and virials. The default values for these parameters were; 2 meV/atom, 0.1 eV/Å and 0.2 eV, respectively. For high-energy systems, such as liquid structures, dense structures, and short-range interactions, these regularisation terms were increased with a factor ten ($5 \sigma$). The full input of the GAP training are available in [LINK TO REPOSITORY TO BE ADDED UPON ACCEPTANCE]. The number of grid points used in the tabulation was 1000 for both two-body and EAM descriptors and for the three-body grid $100 \times 100 \times 100$.

\subsection{Density functional theory calculation details}\label{sec:dft-calc}

All the DFT calculations in this work were performed using the \textsc{VASP} DFT code~\cite{kresse_ab_1993,kresse_ab_1994,kresse_efficiency_1996,kresse_efficient_1996}. The calculations used the PBE GGA exchange-correlation functional~\cite{perdew_generalized_1996} and projector augmented wave (PAW) pseudopotentials (\texttt{Al}, \texttt{Cr\_pv}, \texttt{Cu\_pv}, \texttt{Fe\_sv}, \texttt{Ni}). The plane-wave expansion energy cutoff was 500 eV for all elements. $k$-points were defined using $\Gamma$-centered Monkhorst–Pack grids~\cite{monkhorst_special_1976} with maximum $k$-spacing of 0.15 Å$^{-1}$. Additionally, first order Methfessel-Paxton smearing of 0.1 eV was applied~\cite{methfessel_high-precision_1989}. 
The convergence criterion of energy was set to
$1.0 \times 10^{-6}$ eV. These DFT parameters were chosen as they allowed us to leverage DFT calculations from previous work~\cite{fellman2024fast,byggmastar2022multiscale}.
Furthermore, the calculations applied collinear spin-polarization with initialized ferromagnetic order. The DFT calculations in this work could be at times very challenging, due to the magnetic complexity of the alloys and large system sizes required to represent the HEA compositions. Particularly, the DFT calculations with all elements and over a hundred atoms regularly failed to converge, which posed some issues calculating reference values. The consequences of this is discussed in section~\ref{sec:R}.

\subsection{Training and testing data}

The quality of the training data is critical for the quality of any ML model. The ML potential presented in this work was designed with the goal of making a good general-purpose model with additional considerations for short-range repulsive interactions relevant to radiation damage simulations. The training data consists of a diverse set of 6\,750 DFT-calculated structures of different sizes (235\,361 atoms in total). The training data focused mostly on the FCC lattice structure, while the BCC structures comprised about 13\% of the total amount of the training data. The training data consists of the following types of systems:

\begin{itemize}
\setlength\itemsep{0.1em}
    \item Randomly distorted unit cells (FCC, BCC and HCP) in the case of elemental data.
    \item FCC, BCC and HCP lattices at finite temperatures and different volumes. Initially created by running MD simulations with earlier iterations of the final models.
    \item Systems containing various vacancy and interstitial configurations and amounts (1-3).
    \item Liquid systems at various densities made by MD melting simulations.
    \item \hkl(100), \hkl(110) and \hkl(111) FCC surfaces (elemental data). 
    \item Surface systems containing a few surface layers in a disordered state (elemental data).
    \item Dimers and trimers with various distances between atoms.  
    \item Short-range systems where an atom is randomly inserted into a FCC lattice close to other atoms, but not too close ($\sim>$1 Å) . 
    \item Binary, ternary, quaternary and quinary alloy compositions.
    \item Ordered and intermetallic phases. 
\end{itemize}

% pure element
For the pure element training data we used DFT databases from our previous work in (Al, Cu, Ni)~\cite{fellman2024fast} and Fe~\cite{byggmastar2022multiscale}. From these databases, farthest point sampling using the SOAP descriptor~\cite{bartok_representing_2013,de_comparing_2016} was performed to choose a subset of between 300 and 600 structures, each in a way that it still contained structures of each class (liquids, surfaces, etc.). The structures not chosen were added to the testing set. For the case of elemental Cr a similar sized data set containing similar structures as for the other pure elements was created and split in the same way. 

% compositions
For alloys we created data for all binary, ternary, quaternary combinations. These were created in a way that the compositional space was sparsely uniformly sampled in order to have a wide coverage of different combinations and compositions. In addition to this, approximately 60 compositions were added later with small random distortions to the atom positions and cell shape. Furthermore, dimer and trimer data was created for each atom pair and triplet. Additionally, known ordered or intermetallic structures were added in to the training data.

% HEA
For the full quinary (five-element) composition an extensive number of random compositions were created. Additionally, the compositional space of Al$_x$CrCuFeNi$_y$, where $x$ is 0--1 and $y$ is 1--3 was given extra focus. This range was chosen based on assessment of typical ranges of experimentally relevant compositions. For the HEA compositions several different structure types were created, including finite temperature systems, liquid structures, short-range systems, systems including vacancies and interstitials. Furthermore, during the development of the potential an iterative approach was employed, where earlier versions of the final potential were used in MD simulations and structures from these were added in to the training set. For example, around 50 ordered HEA structures using Monte-Carlo swapping were created in this manner and added to the training data.

Finally, a random subset of structures from all of the structures categories mentioned were added to the testing data, in addition to the single element structures that were not chosen from farthest point sampling. 

\subsection{Monte Carlo swapping MD}

In order to investigate the ordering, segregation and phase stability of the HEAs, Monte Carlo swapping MD (MCMD) simulations were performed. All MD simulations in this work were run using the \textsc{lammps} simulation package~\cite{thompson_lammps_2022}. In Monte Carlo swapping MD (MCMD) simulations, atom swaps between different chemical types are attempted at regular intervals and the probability of a swap is based on the Metropolis criterion \cite{Met53,PhysRevB.85.184203}. The MCMD simulations in this work were run in the $NPT$ ensemble at 300 K for 200 ps, with atom swap attempts every 10 timesteps and 5 swap attempts between each chemical pair. The time step of the MCMD simulations was set to 0.002 ps. We ran MCMD simulations in different compositions starting from both FCC (500 atoms) and BCC (686 atoms) structures. Additionally, an equiatomic polycrystalline alloy sample with around 42\,000 atoms containing four grains was simulated. Due to the increased system size we increased the swapping attempts by a factor of 100 and ran the simulation for 20 ps.

\subsection{Overlapping cascade simulations}\label{sec:methods:overlap}

Massively overlapping cascade simulations \cite{Nord01,granberg_mechanism_2016} were performed with the developed tabGAP potential. The systems contained 256\,000 atoms with the FCC lattice structure and periodic boundary conditions applied in all directions. 400 consecutive collision cascade simulations with 5 keV recoil energy were performed to accumulate a significant dose. Initially the samples were relaxed to 300 K before irradiation. For each individual simulation, the simulation cell was randomly shifted, atoms that crossed the boundaries were returned back to the other side according to the periodic boundaries \cite{Nord01} and a cascade event was initiated in the centre of the cell. This was done in order to uniformly distribute the recoils within the systems. The individual cascade events are initiated by choosing an atom from the middle of the simulation cell and giving it a corresponding kinetic energy in a random direction sampled from a uniform spherical distribution. During the cascade event, electronic stopping was applied on the atoms with a kinetic energy above 10 eV, implemented as a frictional term using precalculated stopping data. The electronic stopping data were computed using SRIM calculations~\cite{ziegler_srim_2010} for each considered composition. 

During cascade simulations, the boundaries of the system were connected to a thermostat ($NVT$) to simulate the dissipation of temperature to the surrounding material. Self-interaction over the periodic boundary was limited by halting and restarting the cascade event if an atom with over 10 eV energy crossed the simulation cell border. The cascade event was simulated for 20 ps, with an additional 10 ps of relaxation ($NPT$) back to the initial 300 K temperature and zero overall pressure. The overlapping cascade simulations used an adaptive time step \cite{Nor94b}, to ensure energy conservation during the high-energy cascades. The final configuration was analysed using the Wigner-Seitz (WS) method~\cite{nordlund_defect_1998} to determine the number of Frenkel pairs produced from the cascade events. Furthermore, DXA~\cite{0965-0393-18-8-085001} analysis was performed, implemented in OVITO~\cite{stukowski2012automated}, in order to find the types and lengths of the dislocations. In order to investigate the differences in clustering we performed a cluster analysis for both interstitial and vacancy clusters, where the cutoff was chosen to be the midpoint between the third and second nearest neighbor distances ($r_{\mathrm{2.5nn}}$). Five different compositions of the Al$_x$CrCuFeNi$_y$ HEA were considered in the overlapping cascades, which are listed in Table~\ref{tab:compositions}. These compositions were chosen to control the Al and Ni contents separately, thereby allowing us to investigate element-specific effects. 

To understand the relationship between electronic structure and composition, we considered valence electron concentration (VEC) and atomic size differences. The VEC is determined by the following equation:
\begin{equation}
    \text{VEC} = \sum_{i=1}^n c_i \cdot \text{VEC}_i,
\end{equation}
where, $c_i$ is the atomic precentage which is multiplied with the VEC$_i$ for that specific element~\cite{guo2011effect}. 
The atomic size difference ($\delta$\%)~\cite{zhang2008solid} is given as:
\begin{equation}
    \delta\% = 100 \sqrt{\sum_{i=1}^n c_i (1 - \frac{r_i}{\overline{r}})^2},
\end{equation}
where $r_i$ is the atomic radius of a specific element $i$ (the radii are taken from Ref.~\cite{senkov2001effect}) and $\overline{r}$ is the weighted average of the atomic radii of all elements. The values of VEC and $\delta\%$ are given in Table \ref{tab:compositions}.

\begin{table}[ht!]
 \caption{Specific compositions (in atom-\%) that we focus on in this work and their corresponding atom percentages. The last two columns show the valence electron concentration and atomic size difference $\delta$\%.}
 \label{tab:compositions}
 \begin{threeparttable}
  \begin{tabular}{l|ccccc|cc}
   \toprule
   Composition & Al & Cr &  Cu & Fe & Ni & VEC & $\delta$\% \\
   \bottomrule
   &&&&&&&\\
    AlCrCuFeNi & 20 & 20 & 20 & 20 & 20 & 7.6 & 5.6\\
    AlCrCuFeNi$_2$ & 16.7 & 16.7 & 16.7 & 16.7 & 33.2 & 8 & 5.3\\
    AlCrCuFeNi$_3$ & 14.3 & 14.3 & 14.3 & 14.3 & 42.8 & 8.3 & 5.0 \\
    Al$_{0.5}$CrCuFeNi & 11.1 & 22.2 & 22.2 & 22.2 & 22.2 & 8.1 & 4.5\\
    CrCuFeNi & 0 & 25 & 25 & 25 & 25 & 8.8 & 1.2\\
   \bottomrule
  \end{tabular}
 \end{threeparttable}
\end{table}

\section{Results}\label{sec:R}

\subsection{Basic properties and validation}\label{sec:R:basic}

To validate the potential and check its performance, several material properties were calculated. However, due to the limited experimental and computational work, direct comparisons are not easy. A standard check of the accuracy (although a fairly limited one) is to calculate the root mean squared (RMS) errors of the testing data not included in the model training. Table \ref{tab:Test_errors} shows the calculated RMS errors for energies and forces. This error is split into crystalline and liquid structures as they have different weights in the training. The liquid structures are relevant to radiation damage simulations and are generally more challenging to describe properly. The short-range structures are not included in these errors as they contain by design very strong forces and high energies that are not, and need not, be trained accurately (the energy RMS errors in this case are of the order of 100 meV/atom). The energy and force RMS errors in Table \ref{tab:Test_errors} demonstrate a good accuracy of the potential, given the complexity of the alloys. 

The computational efficiency of the potential was estimated by running a simple relaxation on a system containing 32\,000 atoms on a single  AMD EPYC 7763 CPU core. The performance (0.22 milliseconds/(atom $\times$ time step)) is about an order of magnitude slower than the single-element tabGAPs which were used for the development of the current potential. This is due to the limits of the cache memory when retrieving the spline coefficients and the larger cutoff radii of the three-body descriptor. In recent years, there has been significant progress in the computational efficiency of ML potentials. In comparison of the current model with the performant implementation of the atomic cluster expansion (PACE) model, which has been shown to be cutting edge in terms of accuracy and efficiency~\cite{lysogorskiy2021performant}, the tabGAP is still roughly 1.5 times faster than the PACE potential for a single element. This performance is excellent considering the complexity of the alloy and, while being somewhat slower compared to single-element tabGAP potentials, the computational efficiency still allows for large-scale simulations (millions of atoms) at reasonable computational cost. Additionally, the computational efficiency of the model could be improved by reducing the three-body descriptors cutoff distance, but based on testing this can have detrimental effects on the accuracy. 

\begin{table}[ht!]
 \caption{Energy and force RMS errors of the testing data for crystalline and liquid structures separately.}
 \label{tab:Test_errors}
 \begin{threeparttable}
  \begin{tabular}{ccc|ccc}
   \toprule
   (meV/atom) & & && (eV/Å) &   \\
   $E_{\mathrm{crystalline}}$ & $E_{\mathrm{liquid}}$ & & & $F_{\mathrm{crystalline}}$ & $F_{\mathrm{liquid}}$ \\
   \bottomrule
   &&&&&\\
    5.6 & 12.6 &&& 0.2 & 0.27\\
   \bottomrule
  \end{tabular}
 \end{threeparttable}
\end{table}

\begin{figure}[ht!]
    \centering
    \includegraphics[width=0.9\linewidth]{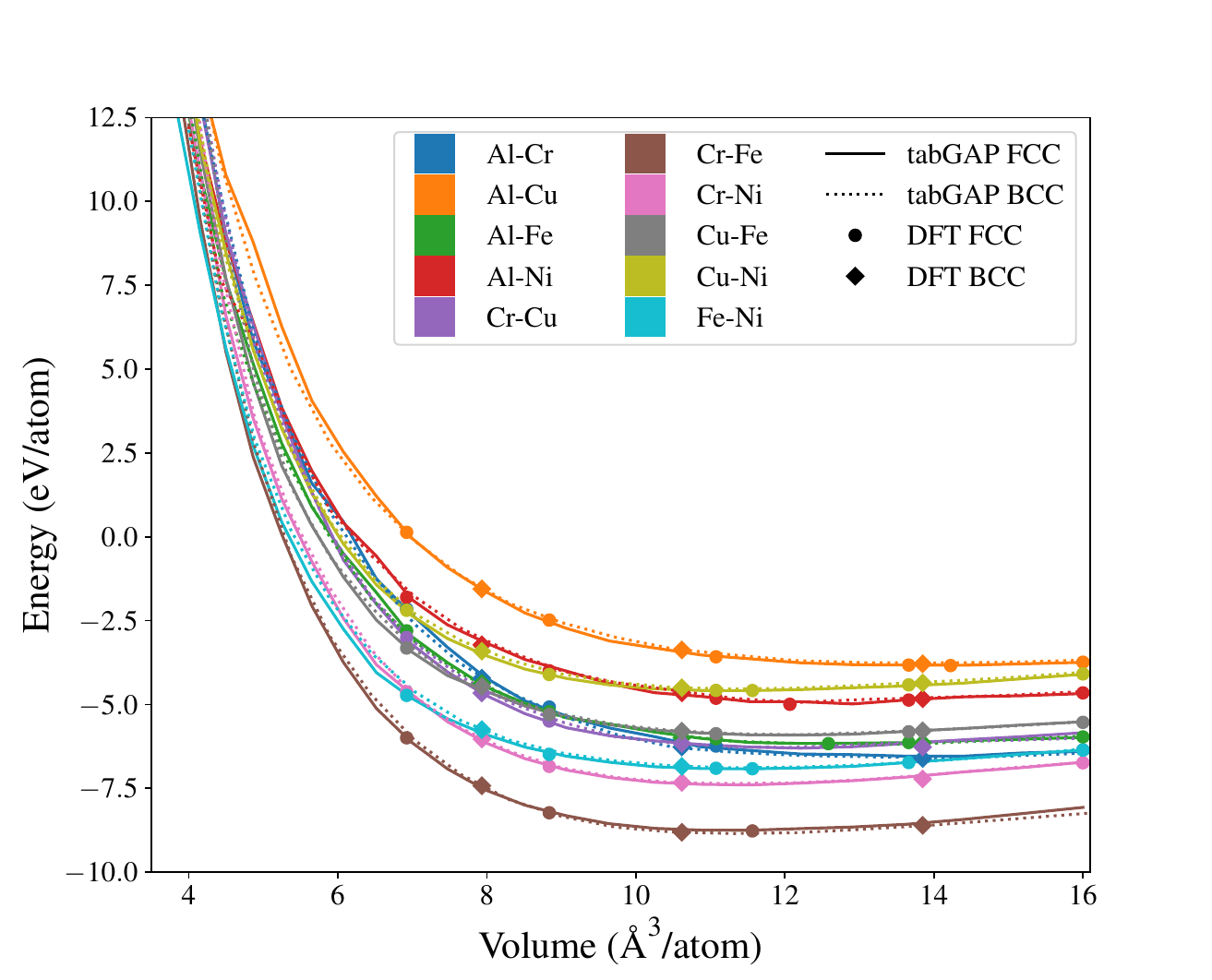}
    \caption{Energy versus volume compared with DFT calculations for all the constituent binary alloys of the HEA at equiatomic concentration and both FCC and BCC phase. Note that the lines are tabGAP results and not plotting lines connecting the dots.}
    \label{fig:Eng-Vol}
\end{figure}

\begin{figure}[ht!]
    \centering
    \includegraphics[width=0.9\linewidth]{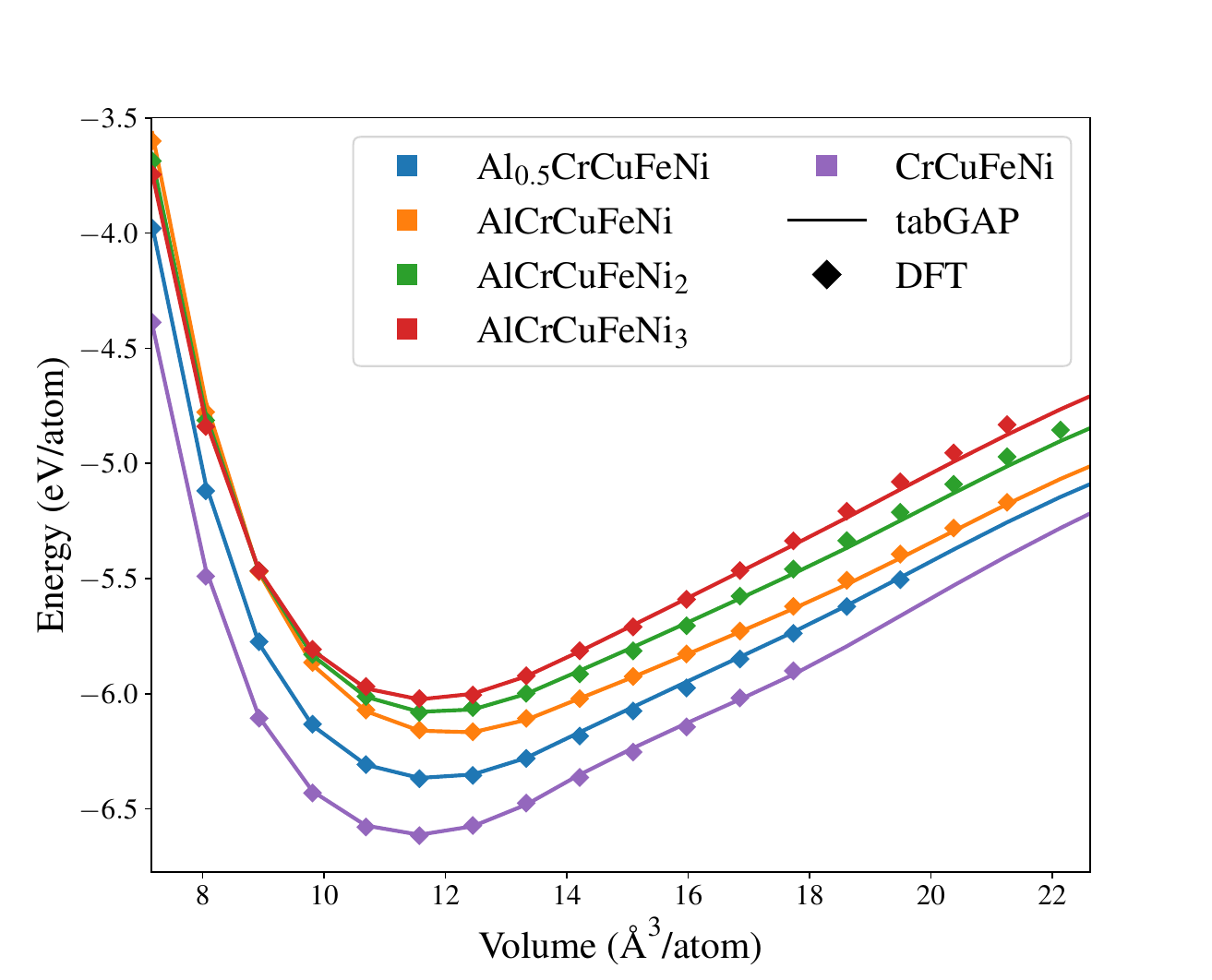}
    \caption{Energy versus volume compared with DFT calculations for the different HEA compositions in the FCC crystal structure, considered in this article.  Note that the lines are tabGAP results and not plotting lines connecting the dots.}
    \label{fig:Eng-Vol-HEA}
\end{figure}

A more representative and physically meaningful validation is the calculation of energy versus volume curves where we can do direct comparison with DFT and check that the potential is smooth and behaves well. Fig.~\ref{fig:Eng-Vol} shows the energy versus volume curves of all the constituent binaries (equiatomic concentration) of the HEA in both FCC and BCC phases compared with DFT calculations. We can see that the potential reproduces DFT with excellent accuracy. In particular we note that the transition towards the external repulsive potential is smooth and well-behaved. Additionally, we can see that both FCC and BCC phases are captured with similar accuracy. Fig.~\ref{fig:Eng-Vol-HEA}, shows the energy-volume curves for all the specific HEA compositions in the FCC crystal structure, considered in this article. In this case, the results are from one random configuration containing 108 atoms in order to get a direct comparison with DFT. Again, we see that the potential shows good agreement with DFT over a wide range of volumes.

\begin{table}[]
    \centering
        \caption{FCC lattice constants $a_0$ (Å) with BCC lattice constant in parentheses, bulk modulus $B$ (GPa), mixing energy (FCC) $E_{\mathrm{mix}}$ (meV/atom), formation energy $E_{\mathrm{form}}$ (meV/atom) and simulated melting point $T_{\mathrm{m}}$ (K). The DFT values are determined from the calculations in Fig.~\ref{fig:Eng-Vol-HEA}. The subscript EOS indicates that these results are calculated using the equations of state.}\label{tab:basic-prop}
    \begin{tabular}{lccc}
         \toprule
        Property & tabGAP & tabGAP$_{\mathrm{EOS}}$ & DFT$_{\mathrm{EOS}}$\\
         \midrule
         AlCrCuFeNi   & &   \\
         \multicolumn{4}{@{}c@{}}{\makebox[\linewidth]{\dashrule[black!40]}} \\[-\jot]
           $a_0$ & 3.643 (2.89) & 3.644 & 3.643\\
           $B$ & - & 148 $\pm$ 5 & 142 \\
           $E_{\mathrm{mix}}$& -128 $\pm$ 3 & -90 $\pm$ 9 & -66 \\
           $E_{\mathrm{form}}^{\mathrm{FCC}}$ & -20 $\pm$ 3 & 20 $\pm$ 9 &  44 \\
           $E_{\mathrm{form}}^{\mathrm{BCC}}$ & -5 $\pm$ 7 & & \\
           $T_{\mathrm{m}}$ & 1400--1500 & & \\
           \midrule
        AlCrCuFeNi$_2$  &  &   \\
        \multicolumn{4}{@{}c@{}}{\makebox[\linewidth]{\dashrule[black!40]}} \\[-\jot]
           $a_0$ & 3.610 (2.87)& 3.619 & 3.620 \\
           $B$ & - & 154 $\pm$ 5 & 151 \\
           $E_{\mathrm{mix}}$& -144 $\pm$ 3 & -115 $\pm$ 7 & -111\\
           $E_{\mathrm{form}}^{\mathrm{FCC}}$& -54 $\pm$ 3 & -25 $\pm$ 7 & -20 \\
           $E_{\mathrm{form}}^{\mathrm{BCC}}$ & -27 $\pm$ 7 & & \\
           $T_{\mathrm{m}}$ & 1440--1500 & &\\
           \midrule
        AlCrCuFeNi$_3$  &  &  \\
        \multicolumn{4}{@{}c@{}}{\makebox[\linewidth]{\dashrule[black!40]}} \\[-\jot]
           $a_0$ & 3.60 (2.86)& 3.60 & 3.60 \\
           $B$ & - & 160 $\pm$ 4 & 159 \\
           $E_{\mathrm{mix}}$& -150 $\pm$ 4 & -123 $\pm$ 10 & -115\\
           $E_{\mathrm{form}}^{\mathrm{FCC}}$& -72 $\pm$ 4 & -44 $\pm$ 10 & -36 \\
           $E_{\mathrm{form}}^{\mathrm{BCC}}$ & -37 $\pm$ 6 & & \\
           $T_{\mathrm{m}}$ & 1540--1600 & &\\
           \midrule
        Al$_{0.5}$CrCuFeNi  & &  \\
        \multicolumn{4}{@{}c@{}}{\makebox[\linewidth]{\dashrule[black!40]}} \\[-\jot]
           $a_0$ & 3.61 (2.87) & 3.612 & 3.61\\
           $B$ & - & 158 $\pm$ 7 & 153 \\
           $E_{\mathrm{mix}}$& -76 $\pm$ 3 & -53 $\pm$ 7 & -40 \\
           $E_{\mathrm{form}}^{\mathrm{FCC}}$& 45 $\pm$ 3 &  63 $\pm$ 7 & 80 \\
           $E_{\mathrm{form}}^{\mathrm{BCC}}$ & 69 $\pm$ 5 & & \\
           $T_{\mathrm{m}}$ & 1540--1600 & &\\
           \midrule
       CrCuFeNi  &  &  \\
       \multicolumn{4}{@{}c@{}}{\makebox[\linewidth]{\dashrule[black!40]}} \\[-\jot]
           $a_0$ & 3.585 (2.85) & 3.582 & 3.58\\
           $B$ & - & 194 $\pm$ 5 & 165 \\
           $E_{\mathrm{mix}}$ & 9 $\pm$ 2 & 19 $\pm$ 3 & 41 \\
           $E_{\mathrm{form}}^{\mathrm{FCC}}$& 144 $\pm$ 2 & 155 $\pm$ 3 & 176 \\
           $E_{\mathrm{form}}^{\mathrm{BCC}}$ & 179 $\pm$ 4 & & \\
           $T_{\mathrm{m}}$ & 1600--1800 & &\\
         \bottomrule
    \end{tabular}
    \label{tab:basic}
\end{table}

Table~\ref{tab:basic-prop} lists some basic material properties calculated for the specific alloy compositions. Here we split the tabGAP results into tabGAP and tabGAP$_{\mathrm{EOS}}$, where the latter refers to results obtained using fitting to the Birch-Murnaghan equation of state (EOS)~\cite{hebbache2004ab}. This separation allows the tabGAP$_{\mathrm{EOS}}$ results to be directly compared with DFT$_{\mathrm{EOS}}$ results, as structure optimization would be required for direct comparison, which unfortunately turned out to be far too computationally expensive to perform in DFT due to the large systems sizes needed. The tabGAP lattice constants were estimated by minimization to zero pressure at 0 K. The lattice constant of the AlCrCuFeNi composition in both BCC and FCC are very close to the experimental values reported in the literature as 2.879 Å and 3.642 Å, respectively~\cite{SHIVAM2023171261}. Likewise, the lattice constants of AlCrCuFeNi$_2$ and AlCrCuFeNi$_3$ matches those reported from experiments~\cite{ma2016strain,luo2020tailored}. 

The bulk moduli of the compositions were estimated by varying the volume of a system (108 atoms) by up to 5 \% from the initial equilibrium lattice constants, and the moduli were extracted from fits to the Birch-Murnaghan equation of state. This procedure was performed for 30 random configurations for each composition. We can use the calculations presented in Fig.~\ref{fig:Eng-Vol-HEA} to give a rough estimate of the bulk modulus in DFT. In this case, the equation of state was fitted to the entire volume range of the DFT points visible in Fig.~\ref{fig:Eng-Vol-HEA}. Overall, the agreement between DFT$_{\mathrm{EOS}}$ and tabGAP$_{\mathrm{EOS}}$ in Tab.~\ref{tab:basic-prop} is good. There are two exceptions to this; first, the mixing energy of AlCrCuFeNi (and CrCuFeNi) is lower in tabGAP than in DFT; second, the bulk modulus of CrCuFeNi is significantly lower in DFT. However, if we estimate the bulk modulus (and mixing energy) with tabGAP in the same manner as for DFT (from Fig.~\ref{fig:Eng-Vol-HEA} that is) we obtain the values (163 GPa and 47 meV/atom for CrCuFeNi) which are close to the DFT ones. The mixing energy of CrCuFeNi in both DFT and tabGAP is positive, which means that the alloy is unstable and would prefer to segregate, at least at low temperatures. Both DFT and tabGAP agree on the relative ordering of the mixing energies between the compositions. Another interesting point is that the formation energy of AlCrCuFeNi is positive when no structure optimization is performed (EOS) and negative when the structure is relaxed. This highlights both the importance of relaxation for optimized lattice distortion and the challenge of making direct comparisons between tabGAP and DFT.  Unsurprisingly, the formation and mixing energies are lower when structure optimization is performed. Generally, the BCC formation energies are higher than the FCC formation energies. However, it should be mentioned that the values presented in Table~\ref{tab:basic-prop} are given for random alloys.

In addition, we have simulated the melting temperatures of the different compositions. The melting temperatures were estimated from simulations using the two-phase method~\cite{Mor94}, with 20 K temperature increments. The melting temperatures are given in Tab.~\ref{tab:basic-prop} as lower and upper estimates, where the lower bound gives the temperature below which we see partial recrystallization, and at the upper bound the entire simulation cell melted. Within the given range, we see solid-liquid coexistence where the system only partially melts during MD time scales (400--800 ps). Hence, we expect this range to roughly correspond to the solidus-liquidus range in the phase diagram. In AlCrCuFeNi$_2$ there have been CALPHAD calculations that predict the presence of the liquid phase to start roughly from 1420 K and complete melting at 1480 K~\cite{liu2016tribological}, which is in excellent agreement with our results. Experimentally, a melting temperature of 1515 K was reported for Al$_{0.5}$CrCuFeNi$_2$~\cite{ng2014phase}, which gives us added confidence that the melting temperatures that our ML potential predicts are reasonable. The melting temperature has also been estimated in previous MD studies, with the mixing of EAM and Morse potentials, as 2440 K for bulk AlCrCuFeNi~\cite{zeng2019thermal}. This is significantly higher than what we report and what would be expected from literature. 

\begin{figure}[ht!]
    \centering
    \includegraphics[width=0.95\linewidth]{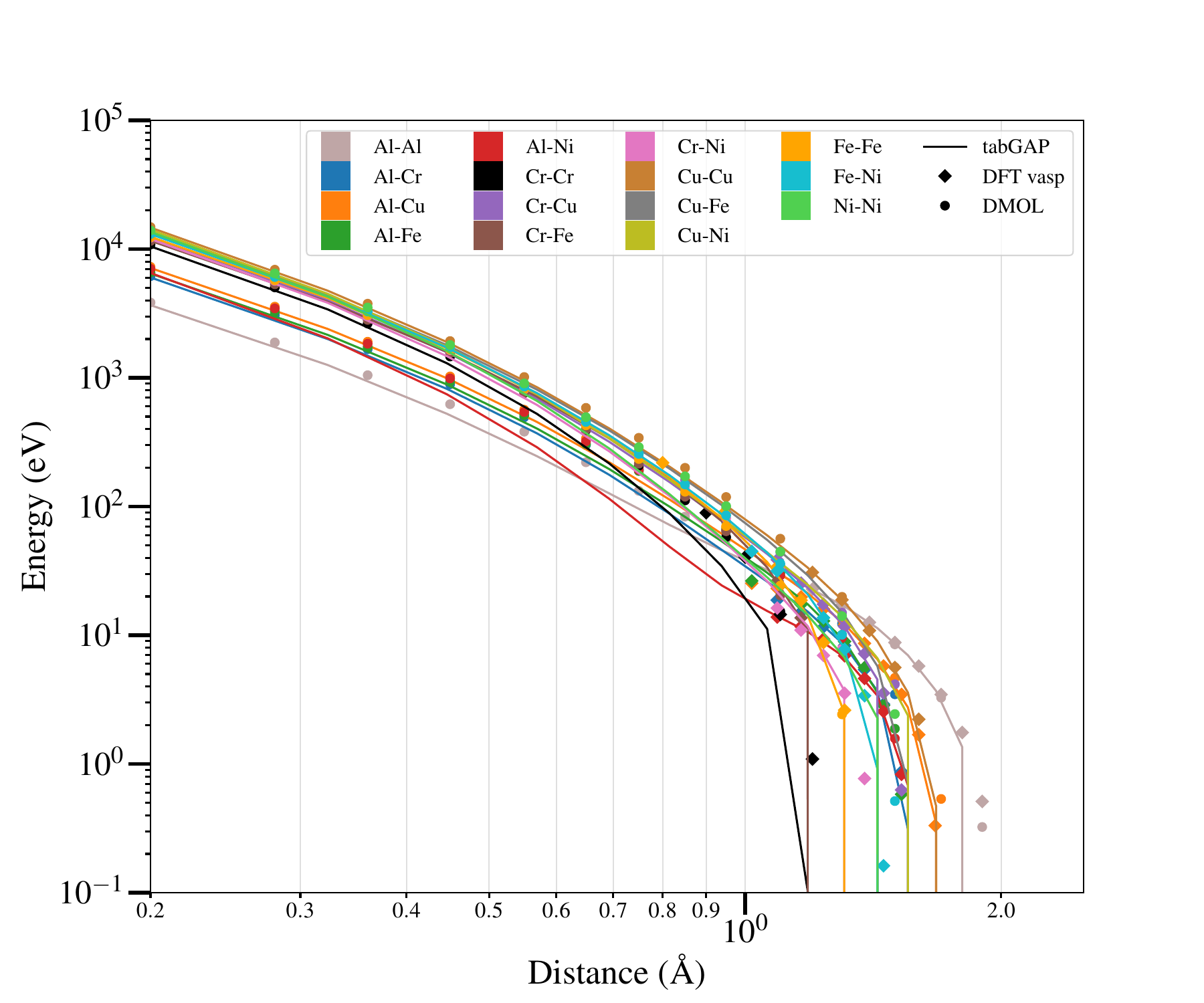}
    \caption{Energy versus distance for all dimer pairs (given in log-log scale) compared with both DFT calculations from VASP and all-electron DFT with DMol~\cite{Nor96c}.}
    \label{fig:dimer}
\end{figure}

Another important property for radiation damage simulations is the repulsive interactions at very short distances. Fig.~\ref{fig:dimer} shows the dimer curves calculated for all dimer pairs in the HEA given in log-log scale. We can see that in the majority of cases the dimer curves transition smoothly from the VASP DFT (this work) to the all-electron DFT data used in the repulsive potentials~\cite{Nor96c}. However, there are some cases (especially Al-Ni and Cr-Cr dimers) that are somewhat ''softer'' than the all-electron DFT at short to medium distances (0.5--1.0 Å). This should be kept in mind when using the potential and interpreting results where high-energy collisions between those pairs are frequent.

\subsection{Mixing energies and vacancy formation energies}

\begin{figure}[ht!]
    \centering
    \includegraphics[trim={0 1cm 0 1cm},clip,width=0.95\linewidth]{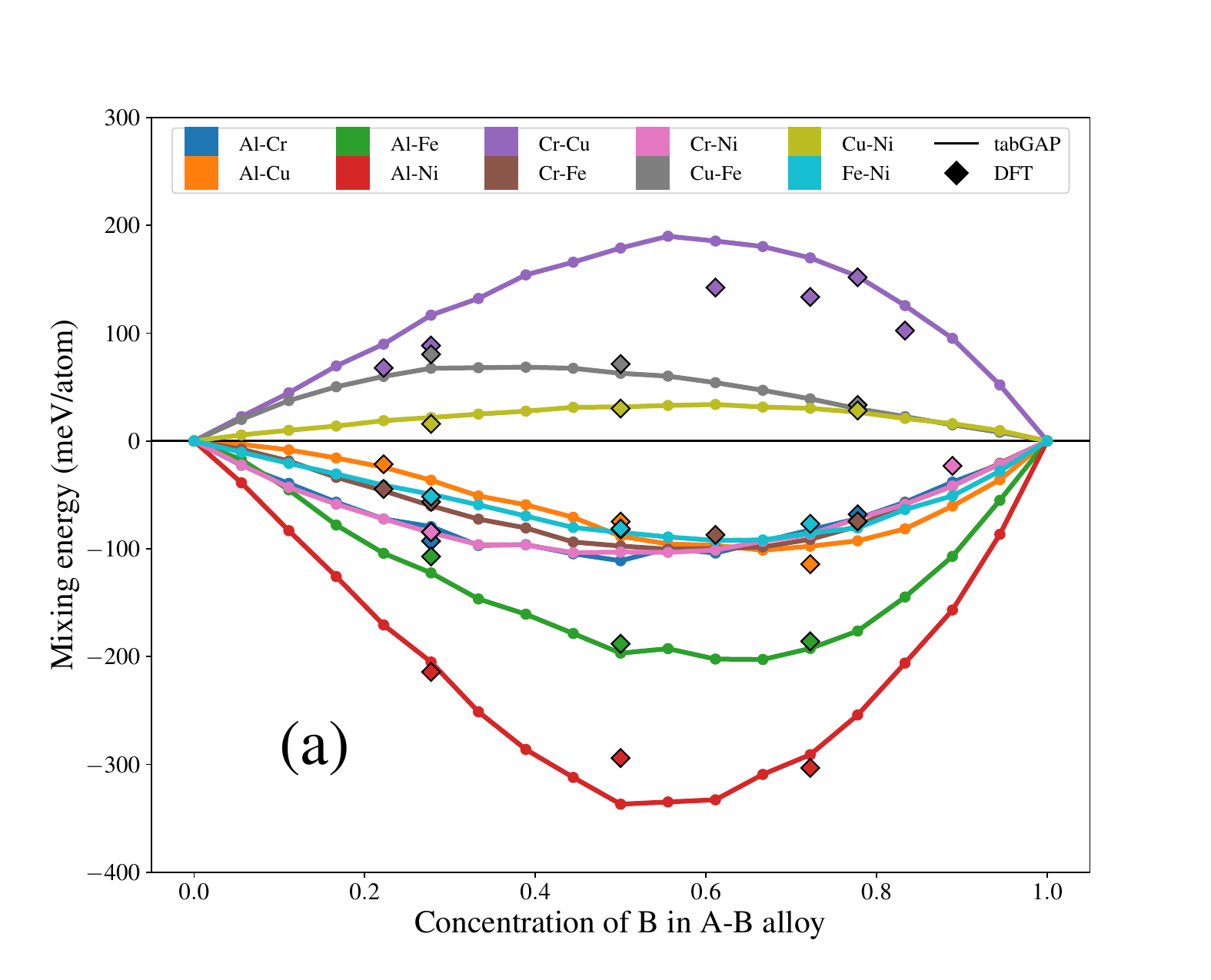}
    \includegraphics[trim={0 1cm 0 1.5cm},clip,width=0.95\linewidth]{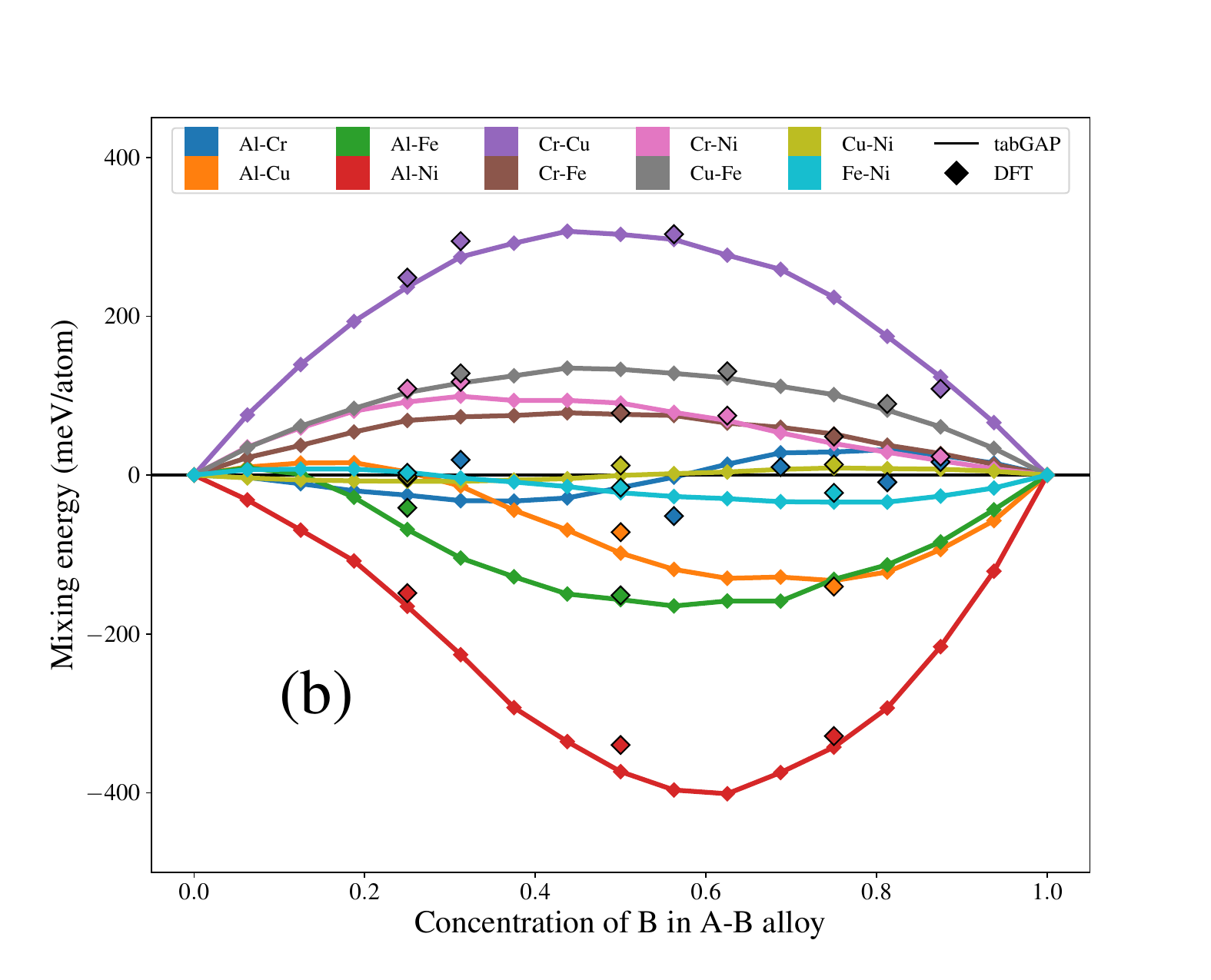}
    \caption{Mixing energies of constituent binary alloys with varying compositions for (a) FCC and (b) BCC. The tabGAP results are averages of five random compositions with the standard deviation being around ($\pm$ 3 meV/atom). } 
    \label{fig:mixing}
\end{figure}

As a further form of validation, the mixing energies of the constituent binaries were also computed. Fig.~\ref{fig:mixing} shows the mixing energies of both FCC and BCC phases. The mixing energies were calculated by creating five random configurations (108 and 128 atoms in FCC and BCC respectively) at each concentration and relaxing the volume of the system and calculating the average mixing energy of the five structures. Structure optimization was not performed on the systems as some of the concentrations are unstable and would spontaneously change lattice structure during minimization. For each case a few DFT points of reference were calculated. These were done by taking some final structures from the MD calculations and calculating their energies using DFT. No DFT structure optimization was performed due to the large system sizes and, therefore, computational cost limitations. There are two notable cases that needs highlighting. Firstly, in FCC the Cr-Cu mixing energy is somewhat overestimated compared to the DFT points. Secondly, in BCC Al-Cr the DFT data points show some disagreement with tabGAP. Considering that the potential was not specifically made for binary alloys the agreement is overall still good. 

\begin{figure}[]
    \centering
    \includegraphics[width=0.95\linewidth]{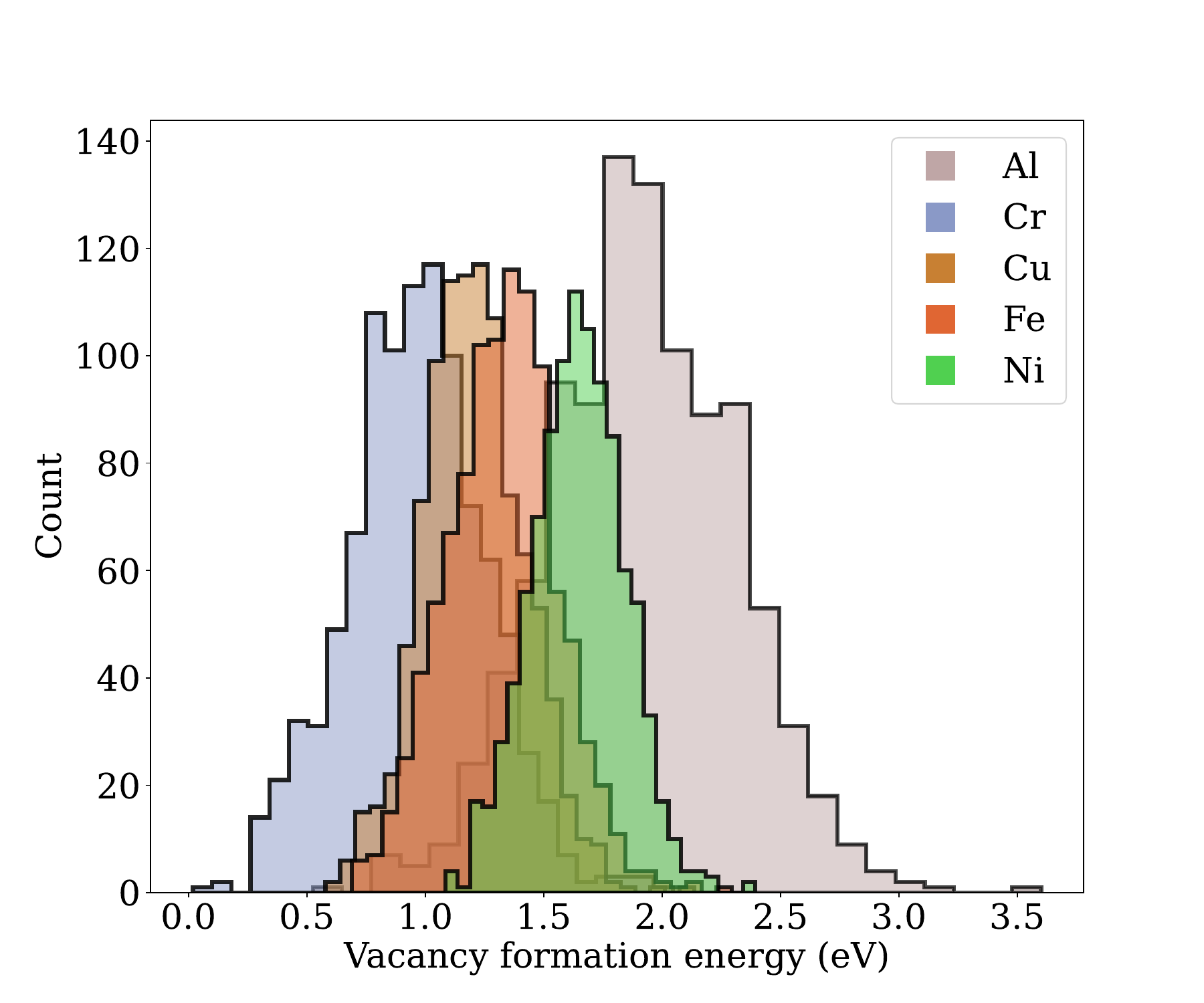}
    \caption{Distribution of vacancy formation energies of different chemical types in AlCrCuFeNi from 1000 random configurations. Number of bins in the histogram was 25. }
    \label{fig:vac-dist}
\end{figure}

\begin{table}[]
    \centering
        \caption{Average vacancy formation energies (eV) of the different chemical species and their standard deviations.}
    \begin{tabular}{lcc}
         \toprule
         composition & element & tabGAP \\
         \midrule
         AlCrCuFeNi & avg. & 1.41 $\pm$ 0.43    \\
         \multicolumn{3}{@{}c@{}}{\makebox[\linewidth]{\dashrule[black!40]}} \\[-\jot]
          & Al & 1.92 $\pm$ 0.39    \\
          & Cr & 0.96 $\pm$ 0.29  \\
          & Cu & 1.20 $\pm$ 0.21  \\
          & Fe & 1.32 $\pm$ 0.23  \\
          & Ni & 1.64 $\pm$ 0.20  \\
           \midrule
        AlCrCuFeNi$_2$ & avg. & 1.45 $\pm$ 0.47    \\
        \multicolumn{3}{@{}c@{}}{\makebox[\linewidth]{\dashrule[black!40]}} \\[-\jot]
          & Al & 2.05 $\pm$ 0.47   \\
          & Cr & 0.99 $\pm$ 0.30  \\
          & Cu & 1.23 $\pm$ 0.23  \\
          & Fe & 1.33 $\pm$ 0.25  \\
          & Ni & 1.63 $\pm$ 0.20 \\
           \midrule
        AlCrCuFeNi$_3$ & avg. & 1.49 $\pm$ 0.49    \\
        \multicolumn{3}{@{}c@{}}{\makebox[\linewidth]{\dashrule[black!40]}} \\[-\jot]
          & Al & 2.19 $\pm$ 0.42   \\
          & Cr & 1.04 $\pm$ 0.31  \\
          & Cu & 1.25 $\pm$ 0.23  \\
          & Fe & 1.37 $\pm$ 0.26  \\
          & Ni & 1.61 $\pm$ 0.20  \\
           \midrule
        Al$_{0.5}$CrCuFeNi & avg. & 1.51 $\pm$ 0.3   \\
        \multicolumn{3}{@{}c@{}}{\makebox[\linewidth]{\dashrule[black!40]}} \\[-\jot]
          & Al & 2.14 $\pm$ 0.35   \\
          & Cr & 1.07 $\pm$ 0.28 \\
          & Cu & 1.25 $\pm$ 0.22 \\
          & Fe & 1.41 $\pm$ 0.22  \\
          & Ni & 1.68 $\pm$ 0.19  \\
           \midrule
       CrCuFeNi & avg. & 1.46 $\pm$ 0.28   \\
       \multicolumn{3}{@{}c@{}}{\makebox[\linewidth]{\dashrule[black!40]}} \\[-\jot]
          & Cr & 1.22 $\pm$ 0.22  \\
          & Cu & 1.33 $\pm$ 0.19  \\
          & Fe & 1.53 $\pm$ 0.20  \\
          & Ni & 1.75 $\pm$ 0.19 \\ 
         \bottomrule
    \end{tabular}
    \label{tab:vac}
\end{table}

Additionally, we computed the formation energies of single vacancies in the HEAs. The vacancy formation energy is calculated as
\begin{equation}
    E_{\mathrm{vac}} = E_{\mathrm{init}} - E_{\mathrm{final}} + \mu_{\mathrm{ref}},
\end{equation}
where $E_{\mathrm{init}}$ is the total energy of the bulk system with a vacancy, $E_{\mathrm{final}}$ is the total energy of the system without a vacancy, and $\mu_{\mathrm{ref}}$ is the chemical potential of the element that occupies the vacancy in the bulk system. Here we use the energy per atom of the given element in its lowest energy crystal structure as the chemical potential. The vacancy formation energies were calculated in all five alloy compositions considered previously. In alloys the chemical surrounding of the vacancy is fluctuating, meaning there is no singular value but a distribution. Therefore, for each composition 1000 random configurations were made and in each random configuration the vacancy formation energy was calculated for each chemical species. Additionally, in the equiatomic composition 13 individual formation energies were calculated with both DFT and tabGAP. This was done to get a crude estimate of the accuracy of tabGAP compared to DFT. The RMS error of the formation energy between DFT and tabGAP is 0.23 eV. While this error is not negligible, we note that the DFT structures were not relaxed due to unreasonable computational cost, which introduces an unknown systematic error. In a ''fair'' comparison (i.e. more statistics and DFT structure optimization) this error could very well change. Additionally, this comparison is rather limited as the chemical environments can vary a lot in complex alloys and a very large amount of DFT calculations would be needed to get statistically significant results. Nevertheless, the qualitative comparison is sufficient to validate that the tabGAP predicts reasonable vacancy formation energies. 

Fig.~\ref{fig:vac-dist} shows a histogram of the distribution of vacancy formation energies in the equiatomic HEA. From the figure we can see that there is a large spread in the vacancy formation energies. Additionally, there is a clear ordering with regard to the chemical species of the removed atom, with Al having the highest average and the largest spread. There are a few outliers visible in the distribution, namely one Al formation energy that is quite high and a couple of Cr values that are really low. The distributions in the other alloy compositions can be found in the Supplementary material (Fig. S3). Table~\ref{tab:vac} shows the average vacancy formation energies for all compositions and for the chemical species. We can see that the average vacancy formation energy decreases when going towards the equiatomic composition. Additionally, the order between the elements remains the same in all compositions. In Al$_x$CoCrFeNi, DFT calculations have been used to investigate the Al dependency of vacancy formation energies~\cite{shu2023ab}. While not directly comparable (due to addition of cobalt, different Al contents, and different chemical potentials used), the vacancy formation energies in this work are somewhat lower than those in Al$_x$CoCrFeNi. However, a similar trend is reported regarding an increase in the Al content, which reduces the vacancy formation energies.

\subsection{Point defect migration energies}

\begin{figure}[]
    \centering
    \includegraphics[width=0.99\linewidth]{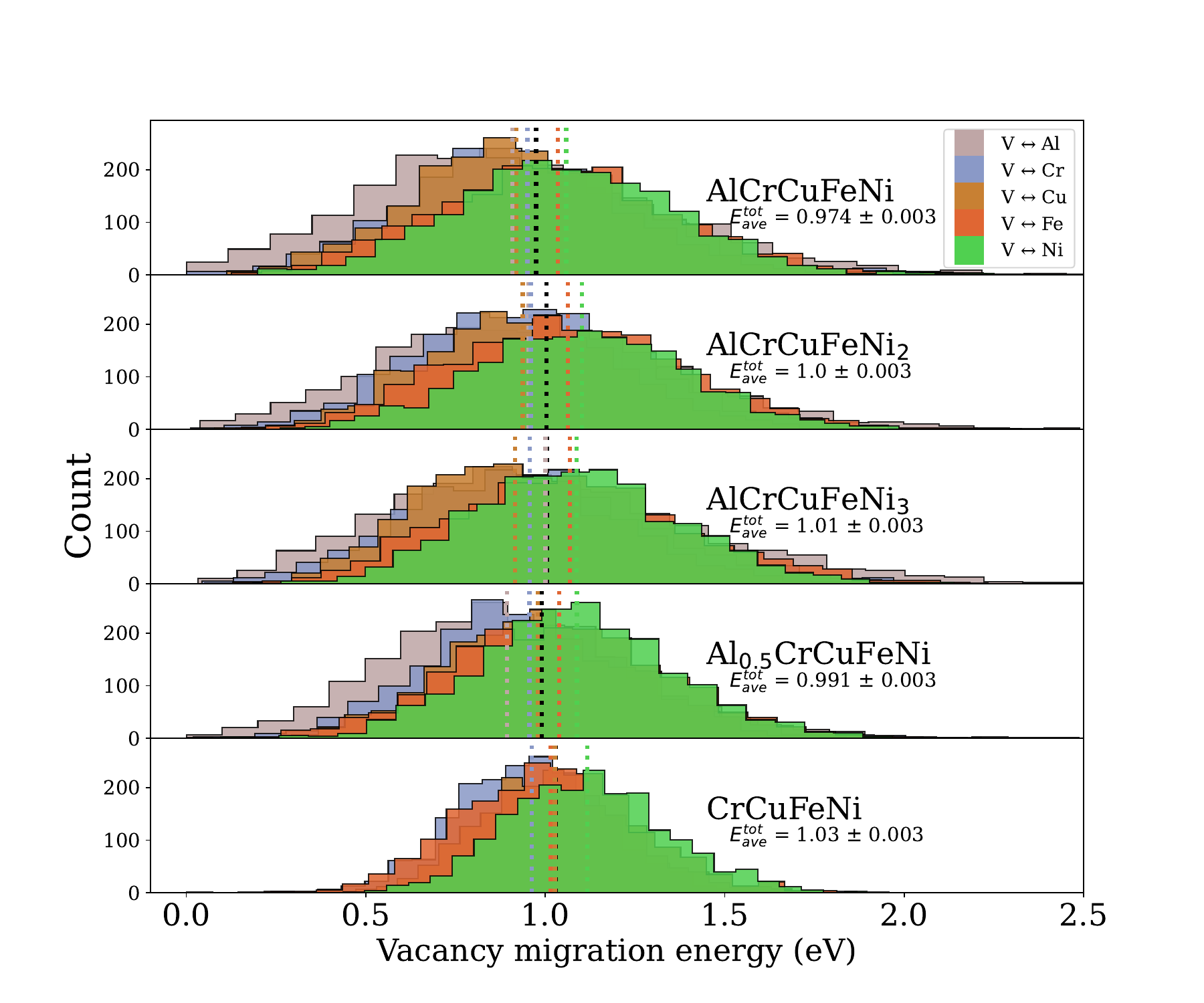}
    \caption{Distributions of vacancy migration energies from 1000 random configurations. The number of bins in the histogram is 25. Vertical dotted lines indicate the average migration energy of exchange with each species as well as the total average (black). The numerical value of the total average is given with the standard error of the mean. }
    \label{fig:vac-mig-dist}
\end{figure}

In order to investigate the differences in diffusion behavior between the compositions, nudged elastic band (NEB)~\cite{henkelman2000a,henkelman2000b} calculations were performed to calculate the vacancy and interstitial migration barriers. For the vacancies, 1000 random configurations were calculated with a vacancy migration barrier determined with regards to exchange with each chemical species. Migration barriers in random alloys are asymmetric, meaning that for each barrier we obtain two migration energies. The convergence criterion for the NEB calculations was set to 0.05 eV/Å. Fig.~\ref{fig:vac-mig-dist} shows the distributions of the vacancy migration energies in each composition. We can see that the increase in Al content decreases the average migration energy slightly. A similar trend has been reported in Al$_x$FeCoCrNi and the migration energies are similar in magnitude~\cite{an2025enhanced}. However, the trend is much smaller in our case. Furthermore, Al is also most often represented in the tails of the distributions, and in general the distributions are wider with increased Al content.

\begin{figure}[]
    \centering
    \includegraphics[width=0.99\linewidth]{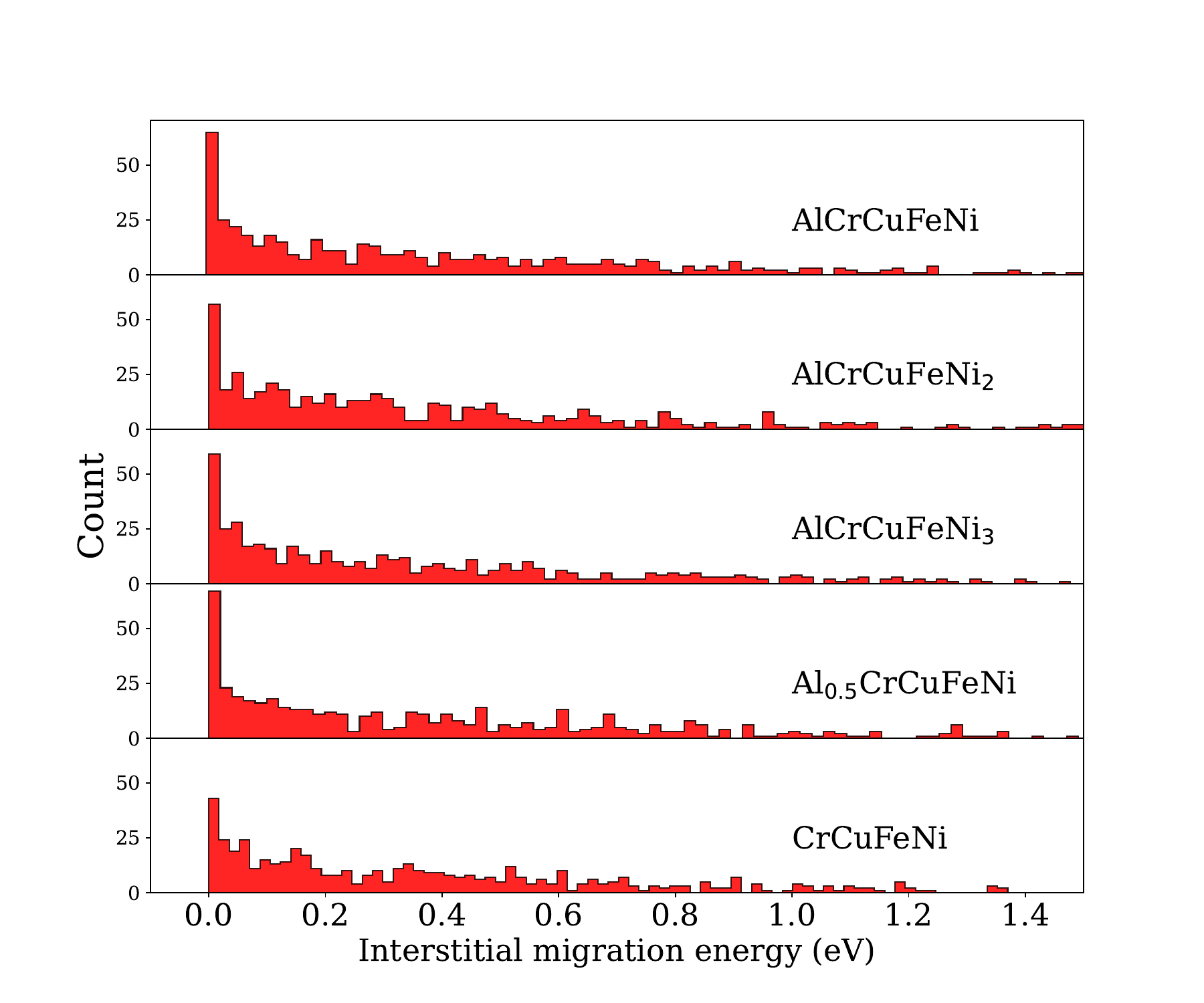}
    \caption{Distributions of interstitial migration energies from 500 individual migration energies. The number of bins in the histogram is 100.}
    \label{fig:int-mig-dist}
\end{figure}

Additionally, we calculated the interstitial migration energies for the different compositions for the \hkl<010> $\rightarrow$ \hkl<001> dumbbell translation-rotation path, which is the lowest-energy path in pure FCC metals. These are shown in Fig.~\ref{fig:int-mig-dist}, where we give the distributions in each composition. In this case, many of the random configurations failed to reach convergence or were unstable. These were omitted, meaning there might be an unknown bias in the distributions due to preferential convergence. The distributions shown are from 500 randomly sampled migration energies from the converged calculations. The possible bias due to convergence and the presence of many high energy outliers makes the estimation of an average migration energy unreliable. In general, the distributions are very similar to each other. Looking at the peak of really low energy barriers, CrCuFeNi stands out from the other compositions, having far fewer of them. Compared to pure elements, where the interstitial migration energies are of the order 0.1 eV ~\cite{fellman2024fast}, the HEA interstitial migration energy distributions includes in all cases far higher energies.

\subsection{FCC/BCC phase stability}

\begin{figure*}[ht!]
    \centering
\begin{subfigure}[t]{0.3\textwidth}
\label{fig:final:Cu}
    \includegraphics[width=\textwidth]{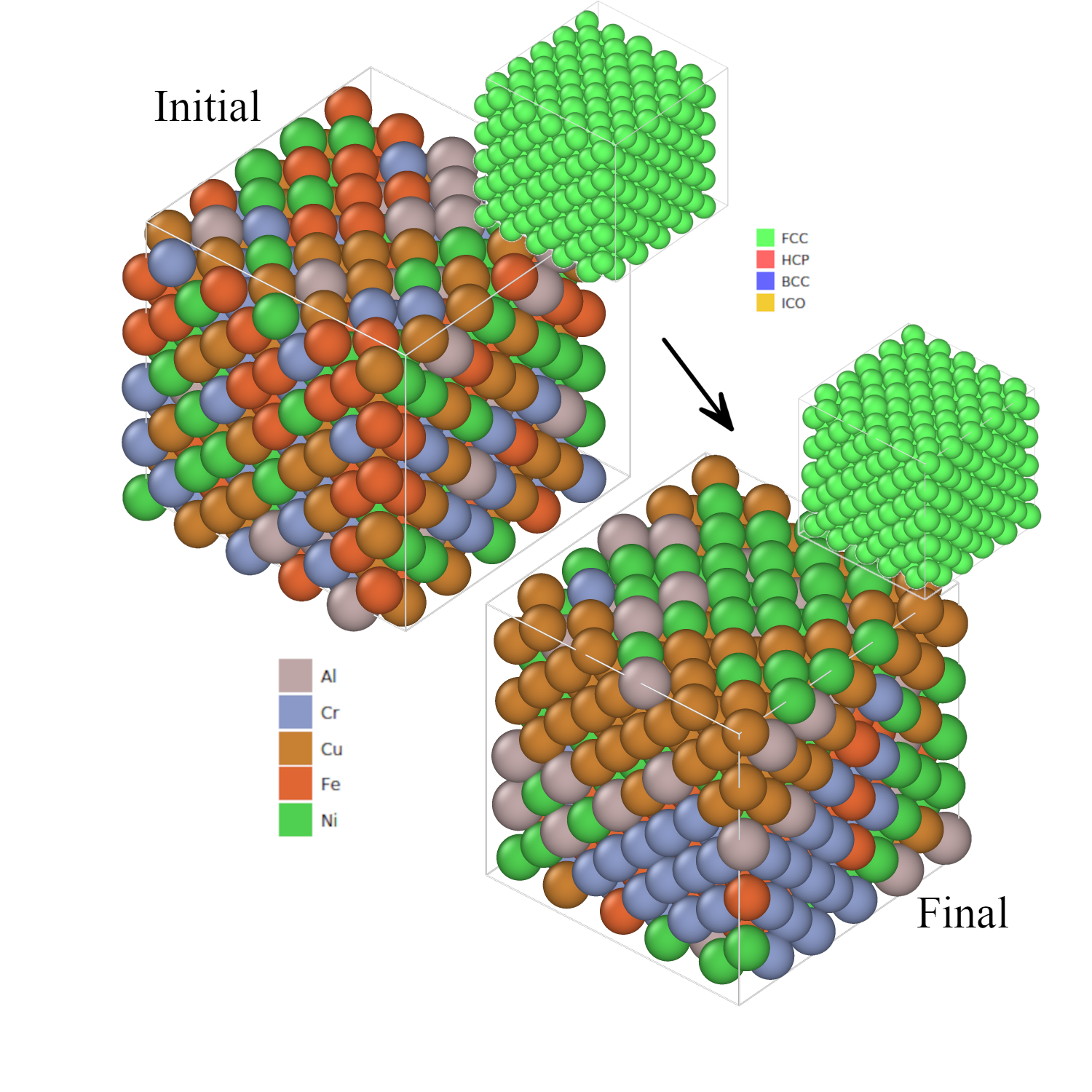}
      \caption{Al$_{0.5}$CrCuFeNi}
\end{subfigure}
\begin{subfigure}[t]{0.3\textwidth}
\label{fig:final:Al}
    \includegraphics[width=\textwidth]{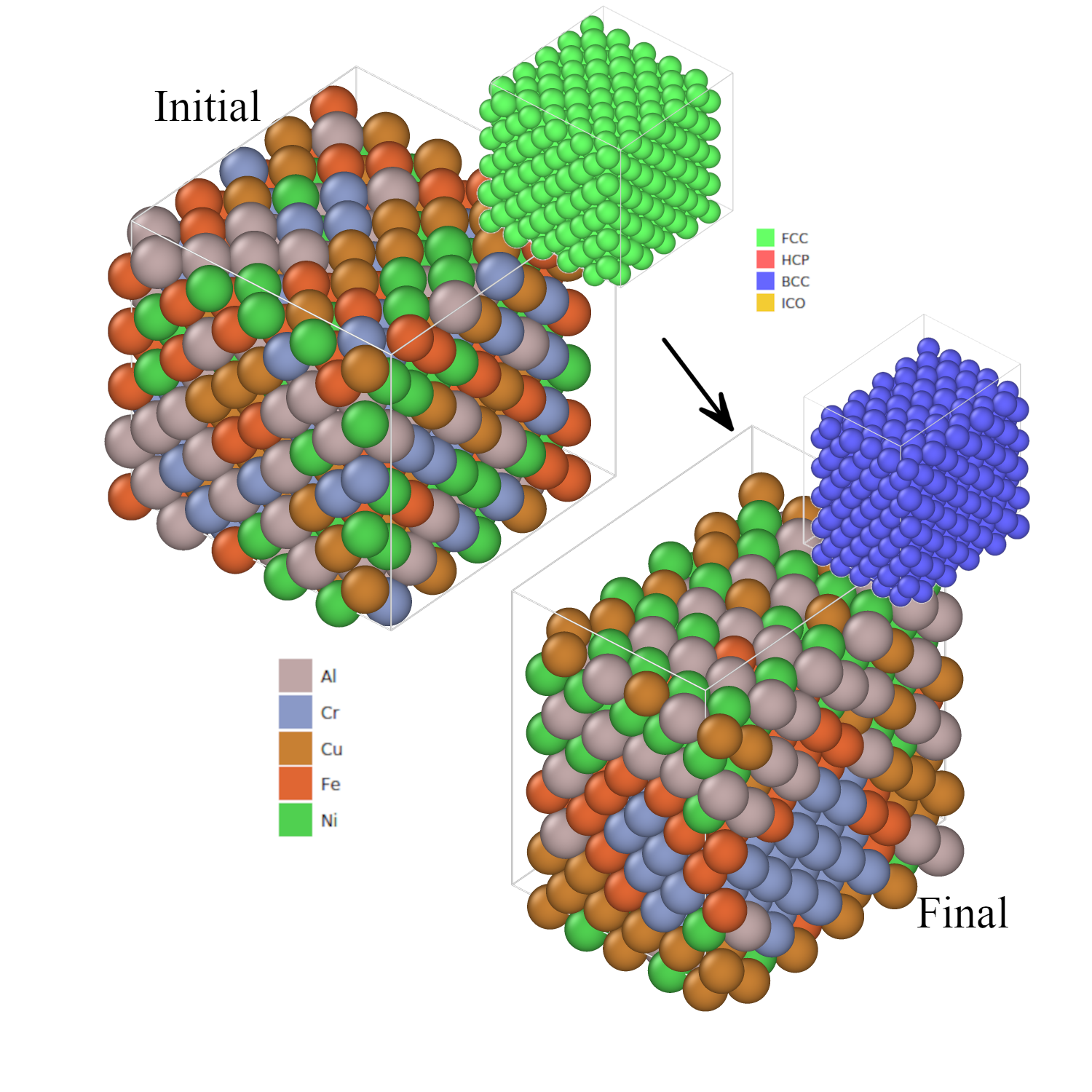}
    \caption{AlCrCuFeNi}
\end{subfigure}
\begin{subfigure}[t]{0.3\textwidth}
\label{fig:final:Ni}
    \includegraphics[width=\textwidth]{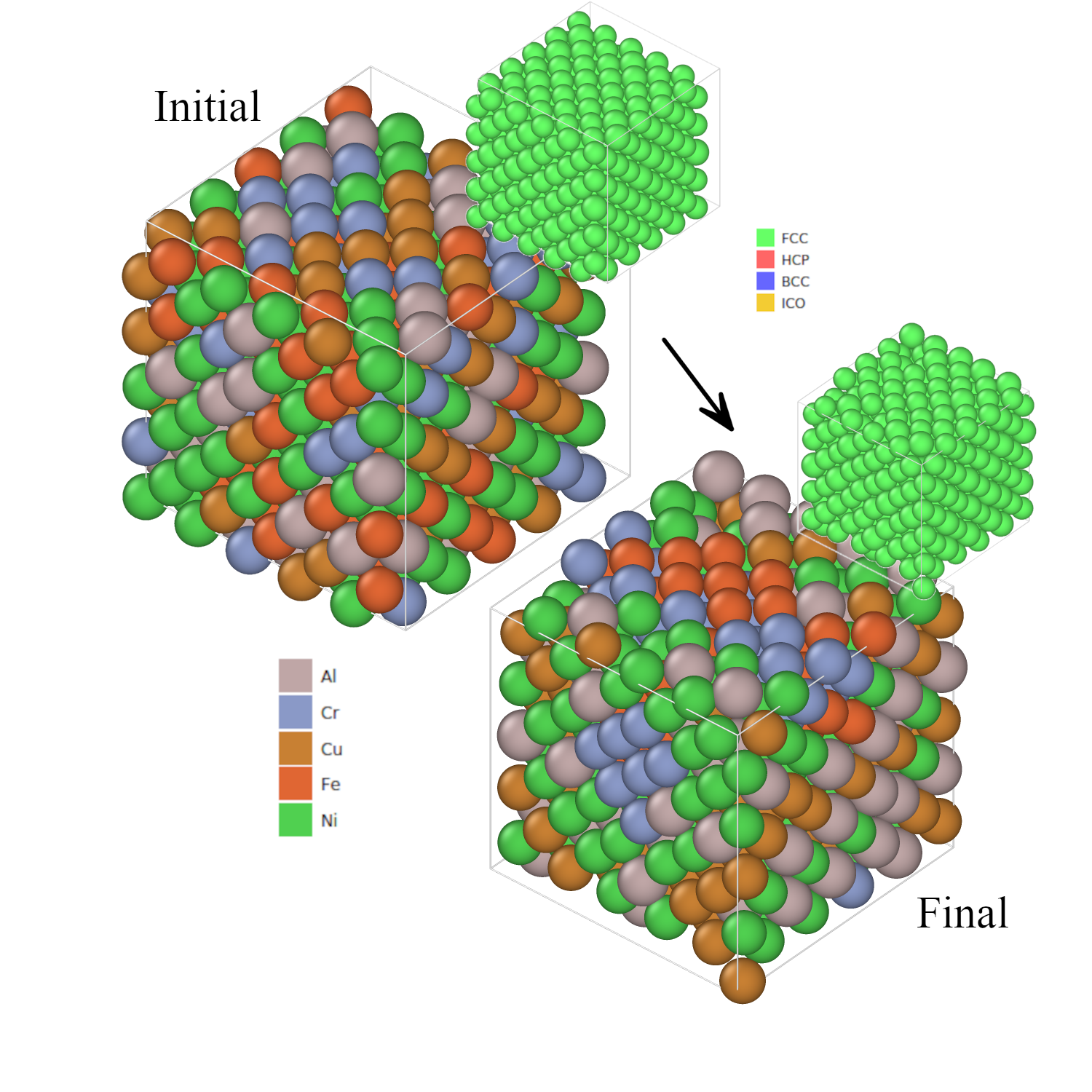}
    \caption{AlCrCuFeNi$_2$}
\end{subfigure}
\begin{subfigure}[t]{0.3\textwidth}
\label{fig:final:Al2}
    \includegraphics[width=\textwidth]{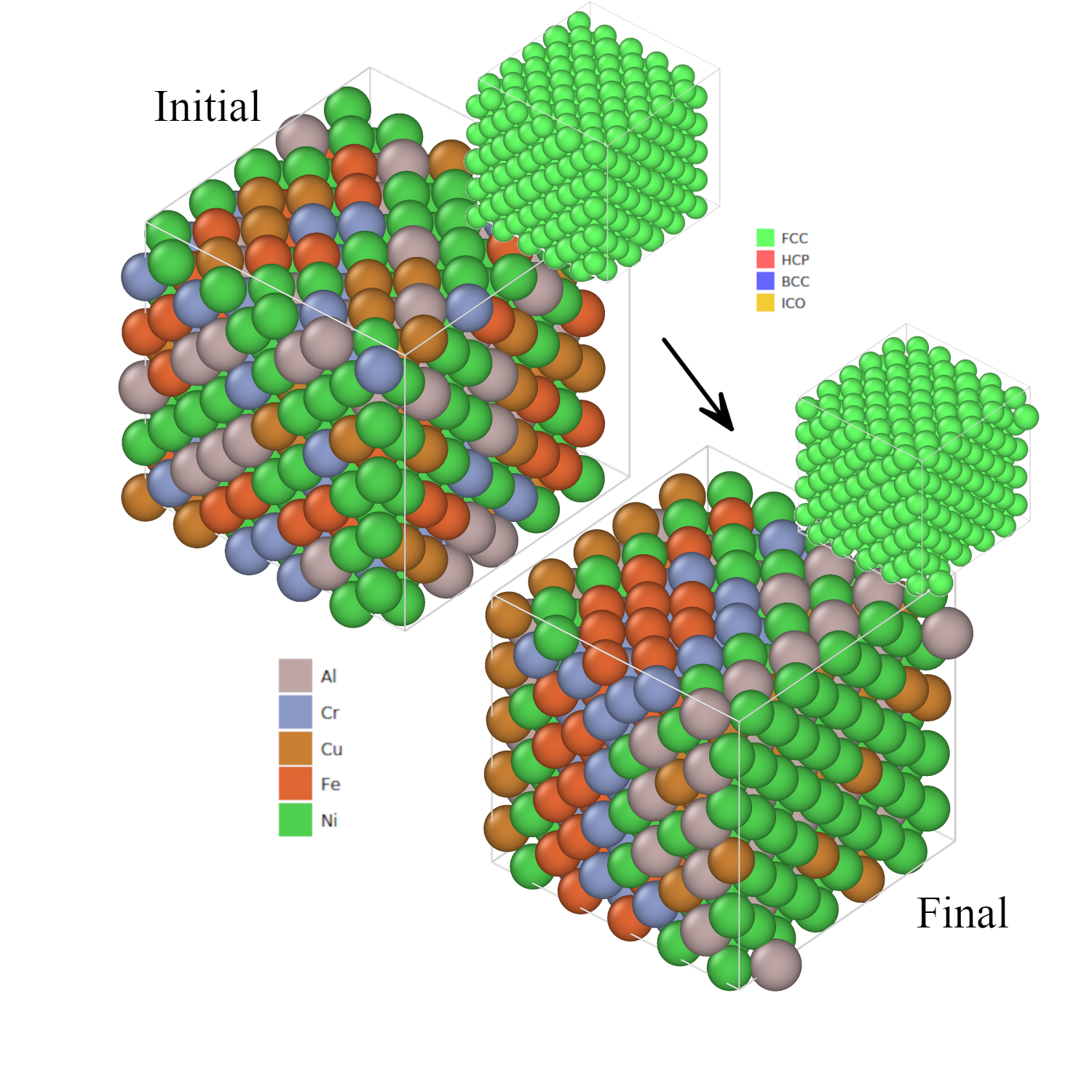}
    \caption{AlCrCuFeNi$_3$}
\end{subfigure}
\begin{subfigure}[t]{0.3\textwidth}
\label{fig:final:Ni2}
    \includegraphics[width=\textwidth]{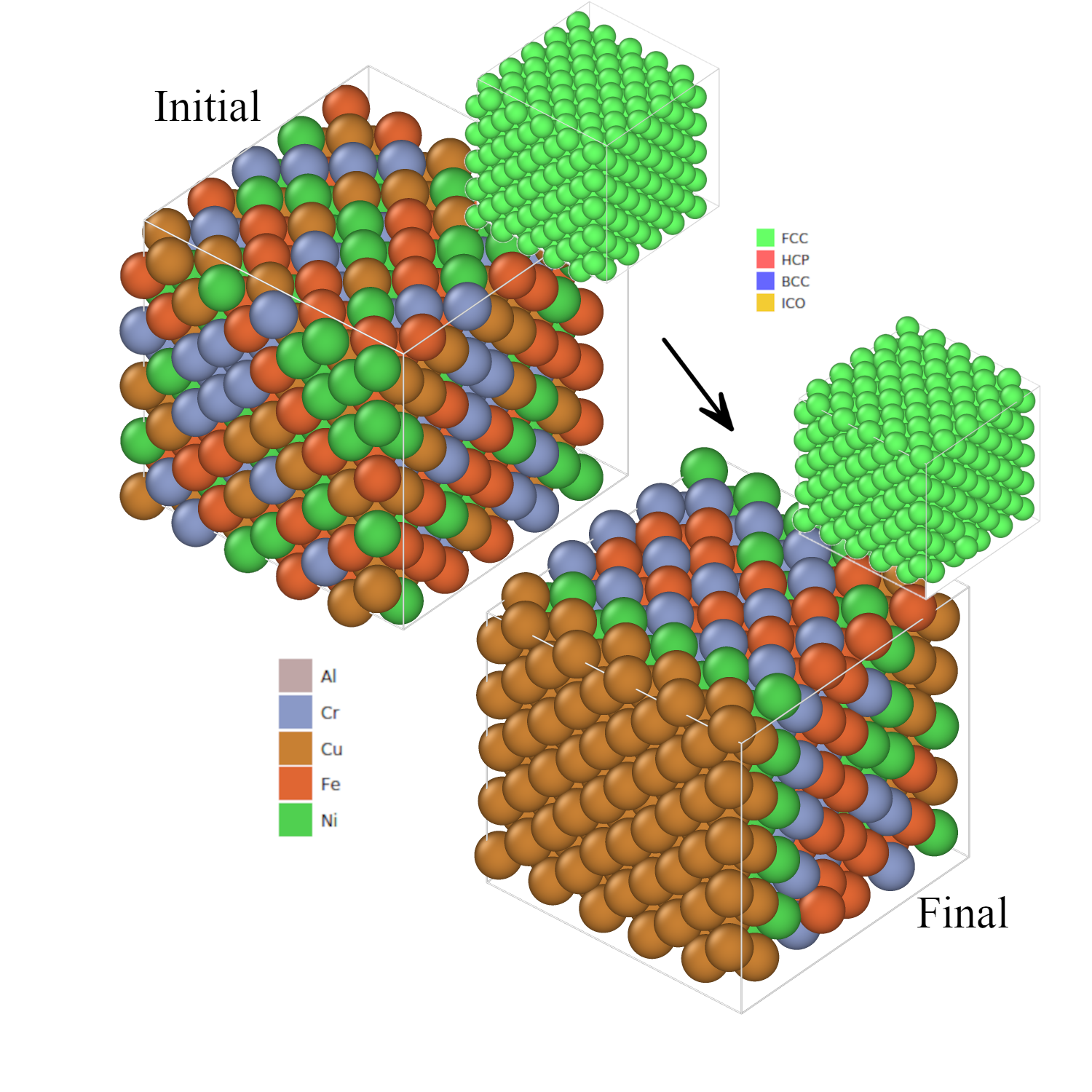}
    \caption{CrCuFeNi}
\end{subfigure}
    \caption{Initial and final frames of Monte Carlo swapping simulations for the five different compositions at 300K. The smaller cells show the crystal structures identified by common neighbor analysis (green FCC and blue BCC).}
    \label{fig:mcmd:small}
\end{figure*}

The stability of the FCC phase for the different alloys is studied in MCMD simulations. Fig.~\ref{fig:mcmd:small} shows the initial and final frames for the different compositions starting initially with an FCC structure. From the snapshots in Fig.~\ref{fig:mcmd:small}, we see the joint segregation of Fe and Cr atoms as well as an additional segregation of Cu. Most interestingly, we observe the change of crystal structure from FCC to BCC for the equiatomic case. In this case, we see a Cr precipitate decorated with Fe and the other elements mixed somewhat uniformly (apart from some Cu segregation). However, when the simulations were performed with larger box sizes (FCC 4\,000 atoms and BCC 2\,000) this transition from FCC to BCC was not observed. The larger systems may be able to accommodate the strain due to nucleation of a new BCC phase easier, and a significantly longer simulation (or more swaps) is needed %or the increased system size stabilizes the alloy and not 
to allow for the formation of large enough nucleation sites of the BCC phase to trigger a phase transition of the entire simulation cell. Therefore, we increased the number of swap attempts for the larger systems so as to match the swap attempts per atom of the small systems and ran the simulations again. With the increased swap attempts we start seeing the formation of the BCC phase in the equiatomic composition, which is inline with what we report for the smaller systems. However, the change to BCC was only partial and there is roughly the same amount of BCC and FCC atoms in the final structures. The final frames of the larger systems can be found in the supplementary materials (Fig. S2). All systems that were initially BCC remained as BCC even after the MCMD simulations. 

In addition to the five specific compositions, we systematically scanned the compositional space of Al$_x$CrCuFeNi$_y$ with the smaller system size by making 100 individual compositions with varying $x$ from 0 to 1 and $y$ from 1 to 3 and looked at the final structures. We observed only the transition from FCC to BCC in a few simulations of compositions close to equiatomic concentrations. However, not every simulation close to the equiatomic composition resulted in phase transition, and in some cases the mixed FCC and BCC phases appeared. In these cases, the BCC phase always consisted of Fe-Cr systems. This suggests that at least one mechanism of the transition from FCC to BCC is the nucleation of a Fe-Cr-rich precipitate (BCC), and if this region is large enough, it may trigger the crystal structure change. Looking at the formation energies in Table~\ref{tab:basic-prop}, we can see that the smallest difference between FCC and BCC is found for the equiatomic composition. In that case, the formation energies are given for a random alloy, and thus the ordering and formation of Fe-Cr precipitates can make the phase change possible due to the small difference in formation energies. We can also see some ordering from the simulations between Fe-Cr and sometimes for Ni-Al. Experimentally, Fe-Cr enrichment and precipitation (Ni-Al as well) have been widely reported~\cite{jinhong2012microstructure,borkar2016hierarchical,guo2017effects,shivam2023microstructural,kafali2023wear}. There are even reports of Fe-Cr-rich BCC nano-precipitates in experimental literature~\cite{luo2021microstructural,su2020microstructure}. Ni-Al also has a strong tendency to order in our simulations, but this combination is more often mixed with other elements, whereas Fe-Cr formed clear precipitates. Similarly, Cu segregation has also been reported for the HEA~\cite{jinhong2012microstructure,guo2013anomalous,borkar2016combinatorial,borkar2016hierarchical,gwalani2017cu,wang2020microstructures,kafali2023wear}. For the equiatomic compositions, the experimental work mostly reports the formation of BCC or coexistence of BCC and FCC with BCC being more prevalent~\cite{wang2020microstructures,rawat2024laser,yadav2025catalytic}. The good agreement between experimental observations and our MCMD simulation results give us confidence that the present AlCrCuFeNi tabGAP potential can provide physical insights into the mechanisms driving phase stability, ordering, and segregation in AlCrCuFeNi alloys. 

\begin{figure*}[ht!]
    \centering
    \includegraphics[width=0.95\linewidth]{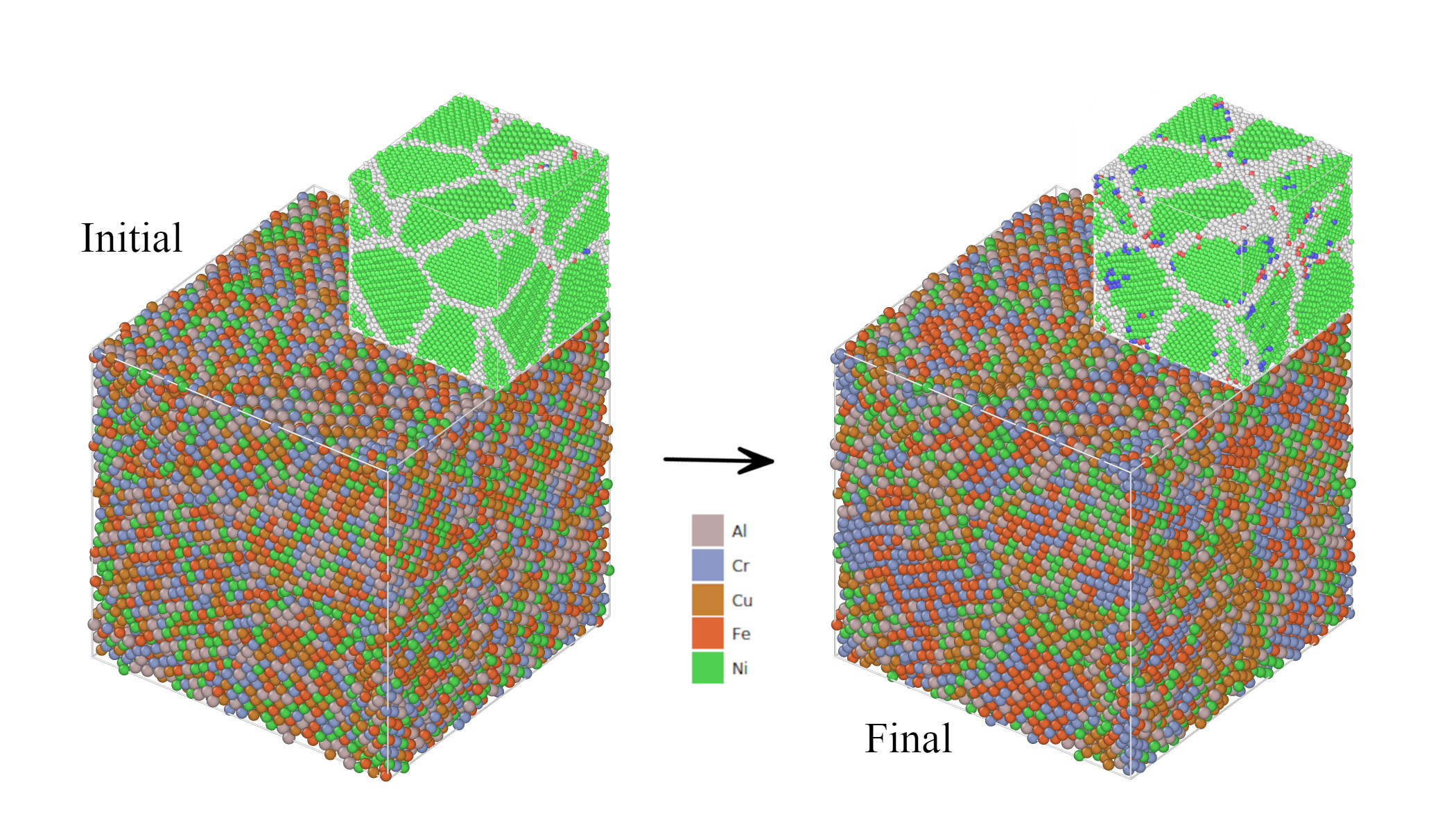}
    \caption{Initial and final frames of Monte Carlo swapping simulations of polycrystalline AlCrCuFeNi at 300 K. The smaller boxes show the crystal structures and grain boundaries from polyhedral template matching analysis, where green color represents FCC atoms, blue represents BCC atoms, red represent HCP atoms and light-gray atoms are of type ''other'' with no identified crystal structure.}
    \label{fig:poly}
\end{figure*}

In order to verify this, additional MCMD simulations of a polycrystalline simulation cell with the equiatomic HEA composition were carried out. Fig.~\ref{fig:poly} shows the initial and final frames of the MCMD simulation of the polycrystalline sample as well as the identified crystal structures from polyhedral template matching (PTM)~\cite{larsen2016robust}. PTM was chosen as it resolves the grain boundary more clearly than common neighbor analysis. From the structures we can see that the system retains its polycrystalline nature, but there are some small regions where BCC crystal structure can be identified in the grains after the simulation close to grain boundaries. Additionally, we see some preferential ordering. In order to give a more quantitative measure of this, we give the compositions of the different crystal structures identified by PTM in table~\ref{tab:polycryst}. We observe that the sample is mostly FCC and ''other'', where the atoms identified as ''other'' are atoms located at the grain boundaries. Atoms identified as BCC or HCP, account for a few percent of all atoms. Most strikingly, we observe that Cr and Fe (especially Fe) is depleted from the grain boundaries. Of the chemical species enriched in the grain boundaries, Cu has the highest fraction, but the differences are not large. Conversely, the BCC atoms mainly consist of Cr, with Fe also making a significant contribution. Additionally, we observe a minor Fe-Cr enrichment in the FCC phase. Experimentally, Cu segregation has been reported at grain boundaries~\cite{wang2020microstructures}. Moreover, Cu-Ni enrichment at grain boundaries has also been reported~\cite{borkar2016combinatorial}. Furthermore, Luo \etal reported that the BCC nano-precipitates have a tendency to pin to grain boundaries~\cite{luo2021microstructural}. These findings are in line with what we observe from MCMD simulations using the presently developed tabGAP potential.

\begin{table}[ht!]
 \caption{The crystal structures identified by polyhedral template matching (fractions in parentheses) in the polycrystalline AlCrCuFeNi system after MCMD simulation and the corresponding atom percentages of each chemical species in the specific crystal structure.}
 \label{tab:polycryst}
 \begin{threeparttable}
  \begin{tabular}{c|ccccc}
   \toprule
   Crystal structure &Al$_{\mathrm{A, \%}}$ & Cr$_{\mathrm{A,\%}}$ &  Cu$_{\mathrm{A, \%}}$ & Fe$_{\mathrm{A, \%}}$ & Ni$_{\mathrm{A, \%}}$ \\
   \bottomrule
   &&&&&\\
    FCC (64.6 \%) & 18.5 & 20.1 & 17.5 & 25.2 & 18.6 \\
    BCC (2.8 \%) & 9.6 & 54.5 & 4.9 & 24.4 & 6.5\\
    HCP (1.8 \%) & 20.9 & 14.9 & 31.0 & 11.7 & 21.5\\
    Other (30.8 \%) & 23.9 & 17.0 & 25.9 & 9.1 & 24.1 \\
   \bottomrule
  \end{tabular}
 \end{threeparttable}
\end{table}

\subsection{Radiation damage}

\begin{figure}[ht!]
    \centering
    \includegraphics[width=0.9\linewidth]{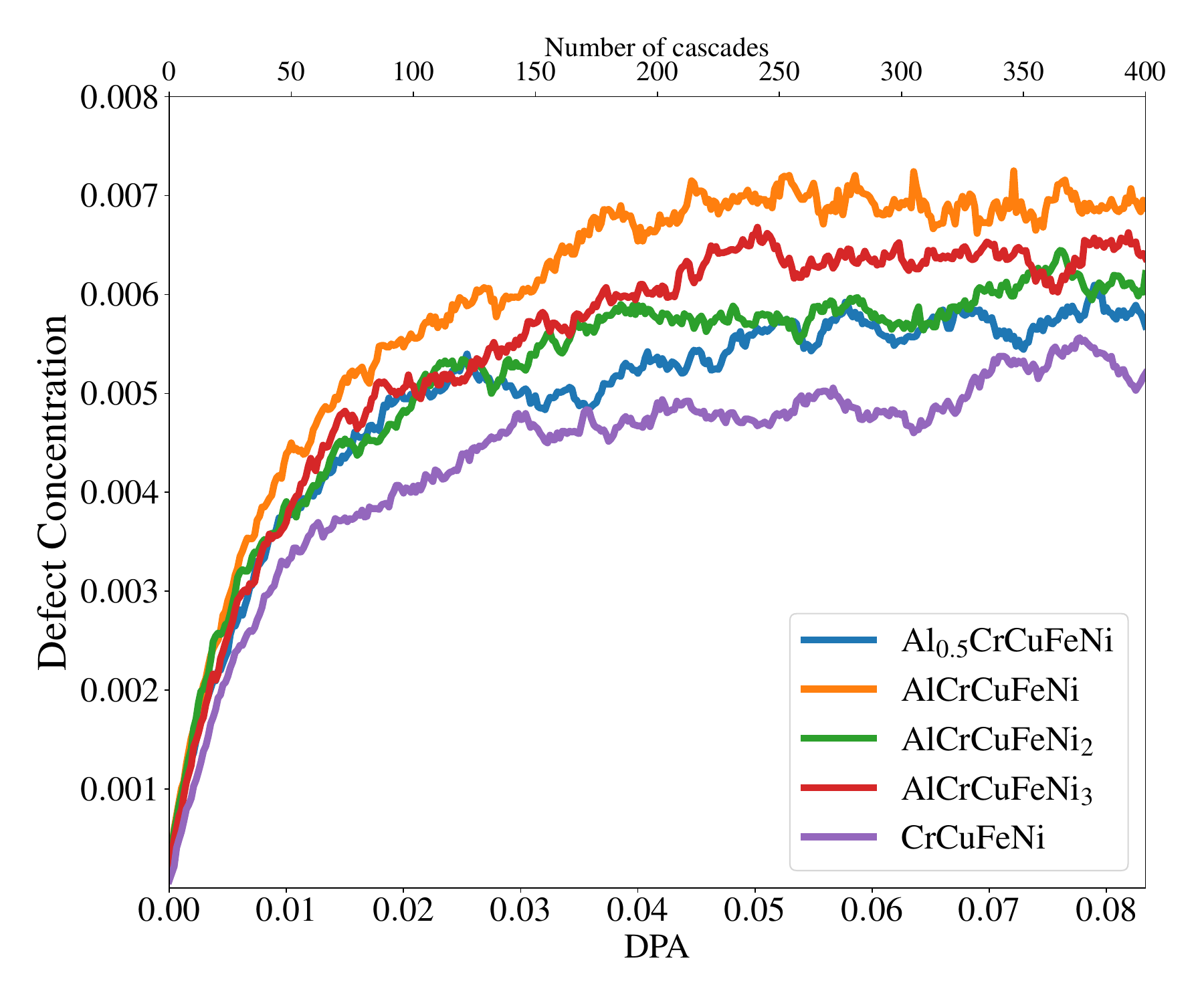}
    \caption{Defect concentration as a function of estimated dose (NRT dpa) for the different compositions. The defect concentration is defined based on Wigner-Seitz analysis including both vacancies and interstitials.}
    \label{fig:defect-concetration}
\end{figure}

\begin{figure*}
    \centering
\begin{subfigure}[t]{0.3\textwidth}
    \includegraphics[width=\textwidth]{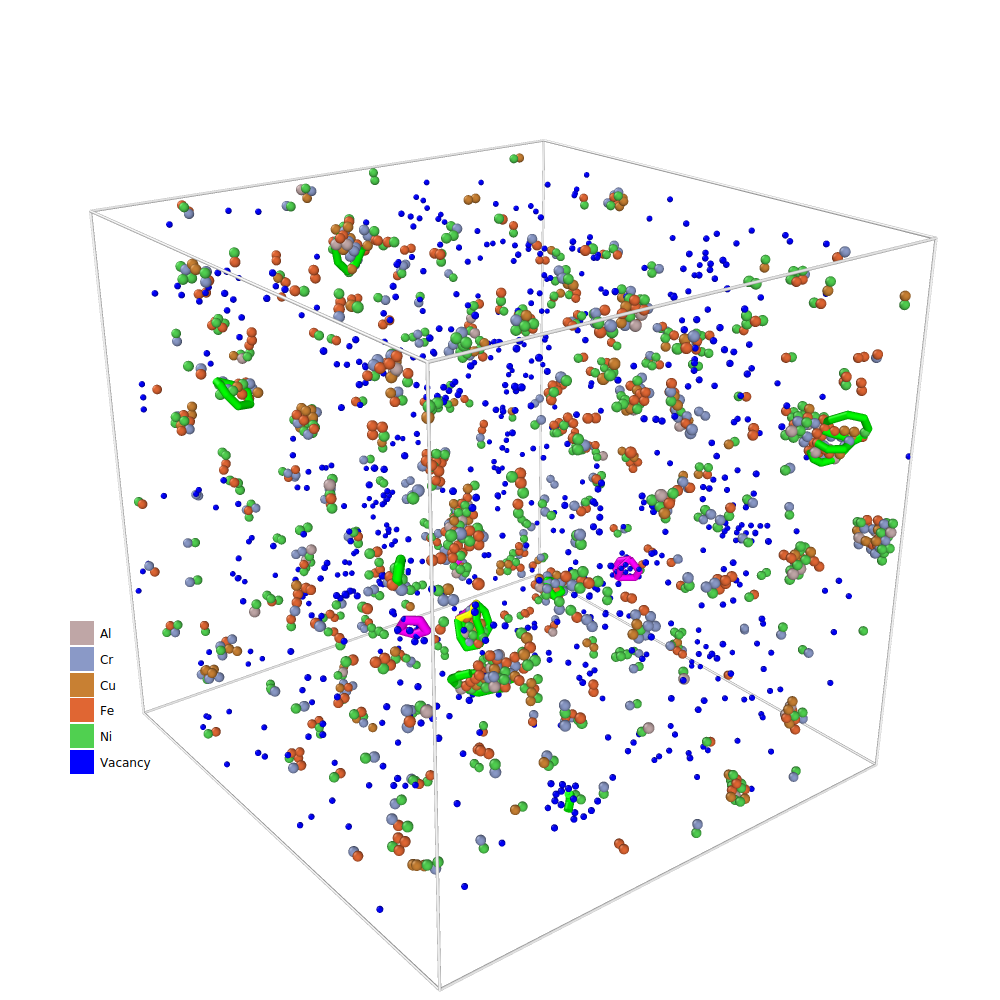}
      \caption{Al$_{0.5}$CrCuFeNi}
\end{subfigure}
\begin{subfigure}[t]{0.3\textwidth}
    \includegraphics[width=\textwidth]{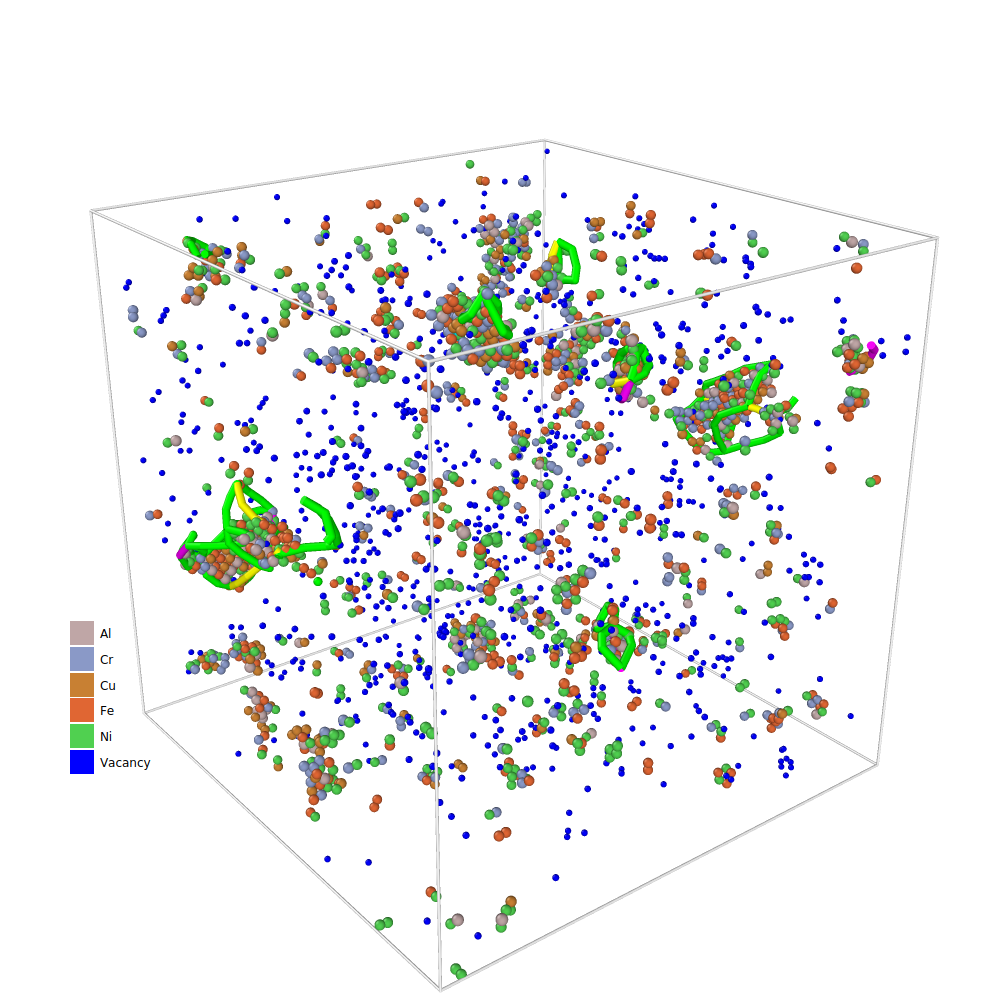}
    \caption{AlCrCuFeNi}
\end{subfigure}
\begin{subfigure}[t]{0.3\textwidth}
    \includegraphics[width=\textwidth]{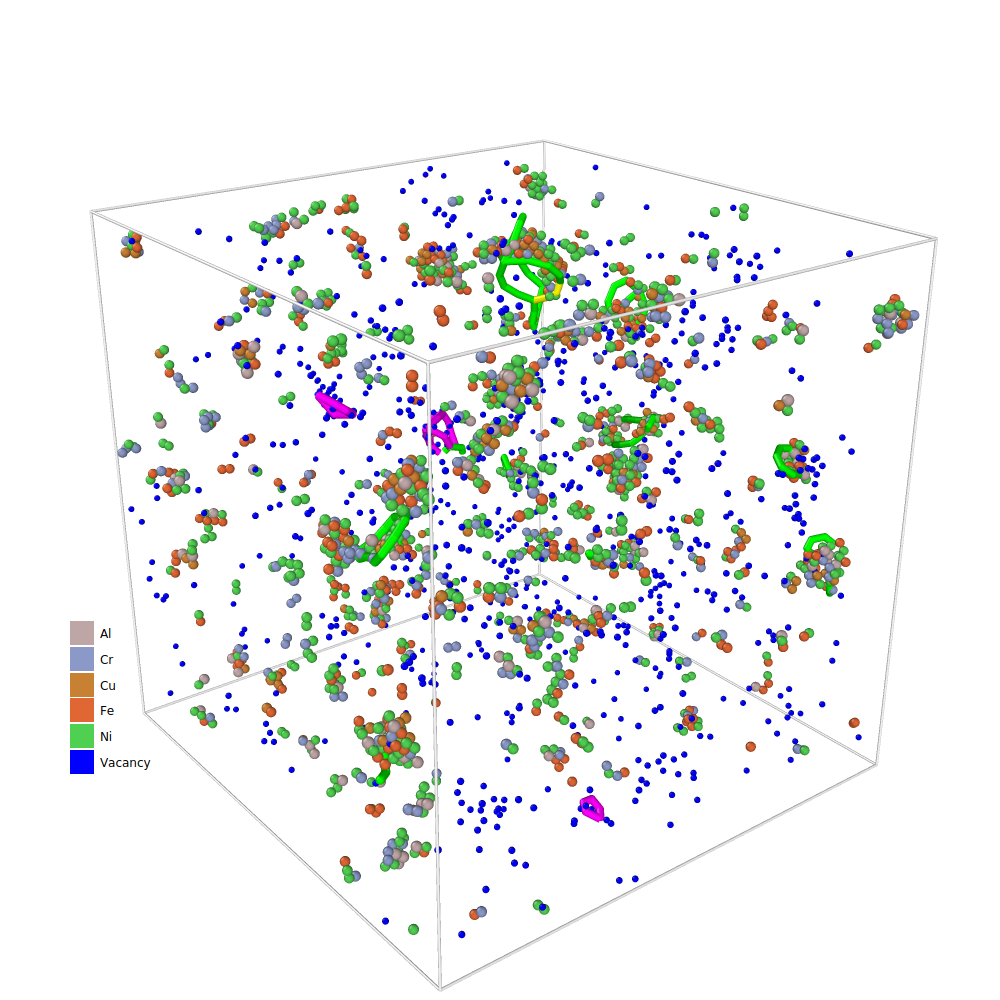}
    \caption{AlCrCuFeNi$_2$}
\end{subfigure}
\begin{subfigure}[t]{0.3\textwidth}
    \includegraphics[width=\textwidth]{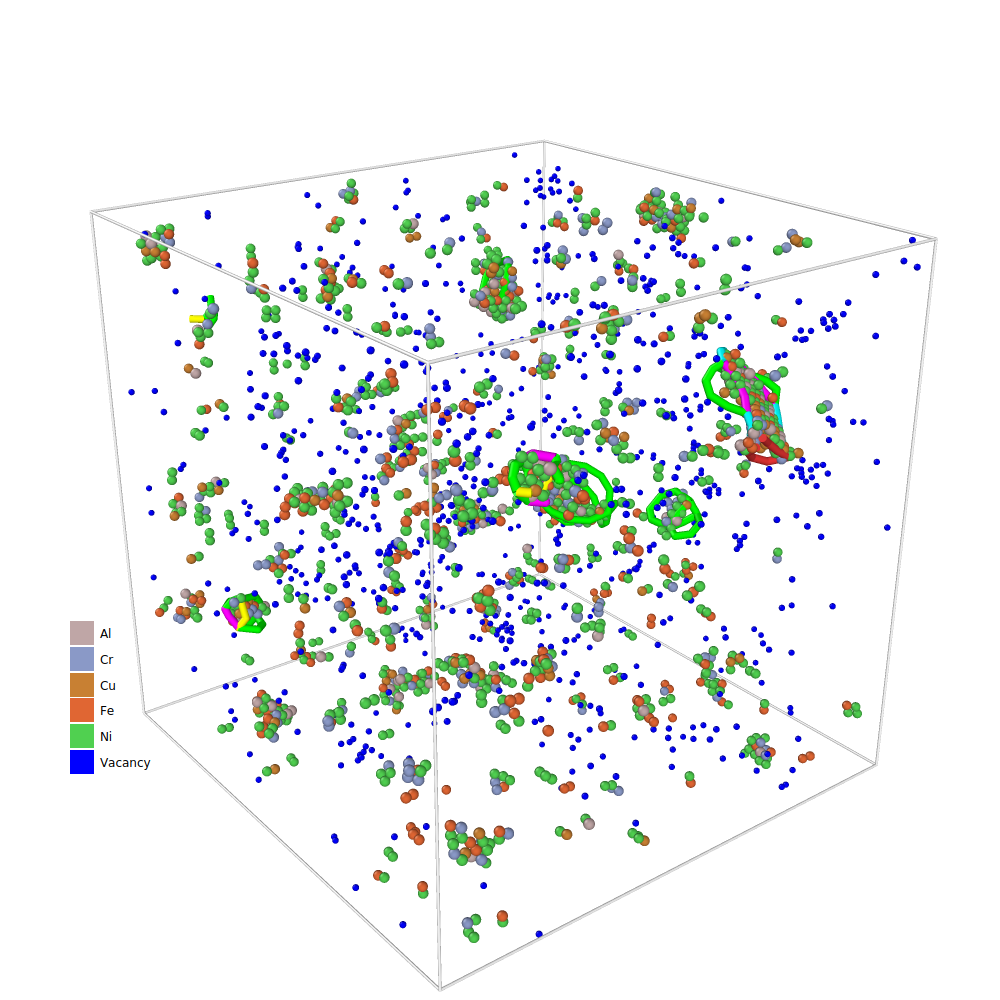}
    \caption{AlCrCuFeNi$_3$}
\end{subfigure}
\begin{subfigure}[t]{0.3\textwidth}
    \includegraphics[width=\textwidth]{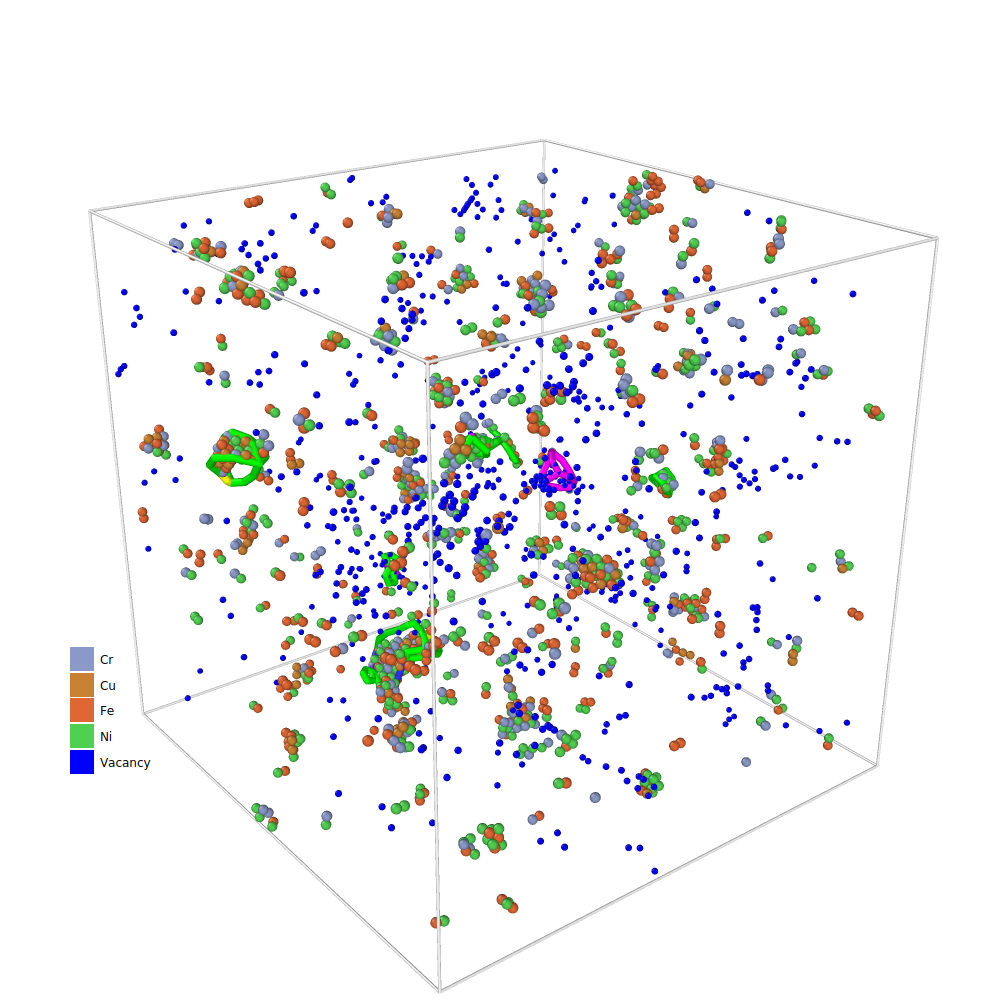}
    \caption{CrCuFeNi}
\end{subfigure}
    \caption{Final configurations after 400 consecutive 5 keV collision cascades in the different compositions. Different colored spheres indicate interstitials of different chemical elements and blue spheres indicate vacancies. Additionally, dislocation lines identified from DXA analysis are shown: Green lines represent Shockley dislocations, cyan lines represent Frank dislocations, yellow lines represent Hirth dislocations, Blue lines represent perfect dislocations, pink lines represent stair-rod dislocations and red lines are dislocations of type ''other''. }
    \label{fig:final:frames}
\end{figure*}

Massively overlapping collision cascade simulations were performed for the five alloy compositions to investigate the dependence of the Al and Ni content on the response of HEA to irradiation. Fig.~\ref{fig:defect-concetration} shows the defect concentration as a function of irradiation dose for the different alloy compositions. The dose was calculated using the NRT-dpa equation~\cite{nordlund_primary_2018}, with an assumed 30 eV threshold displacement energy. 

Initially, we observe the formation of primary damage that is uniformly distributed in the simulation cell. This is indicated by the sharp linear increase in the defect concentration at lower doses. Subsequently, the cascades start to increasingly overlap with the pre-existing defects from the previous cascades, resulting in partial annealing of existing defects and effectively decreasing the growth rate of defects in the irradiated structure. In later stages, the clusters begin to grow and form larger clusters and dislocation structures. In Fig.~\ref{fig:defect-concetration} we see that the defect concentration saturates at $\approx$ 0.06 dpa (300 cascades) %we can see saturation of the defect concentration 
in all of the HEA compositions. 

Fig.~\ref{fig:defect-concetration} reveals significant differences in defect concentrations for different compositions. The highest concentration of the defects we obtained in the equiatomic HEA. %having the highest defect concentration. 
We notice a clear trend with regard to the Al content in the HEA, which can be seen from the evolution of the defect concentrations. Reduction of the Al content, while keeping others proportional, reduces the defect concentration. Surprisingly, we do not observe any clear trend while changing the concentration of Ni. %In regards to Ni concentration, there are clear differences, but no trend is found as in the case of Al. 
Interestingly, such a clear Al dependence of the defect concentration was not observed in recent work on AlCoCrFeNi alloys, where classical potentials were used~\cite{ma2024effect}. The question of whether this is an effect of the used potential or there is a fundamental effect due to %difference when substituting 
substitution of cobalt by copper is interesting but difficult to address. The final defects are visualized in Fig.~\ref{fig:final:frames} for all HEA compositions. Qualitatively, the defect structures are quite similar for all different compositions. In all cases, we see interstitials forming clusters of various sizes with some single interstitial dumbbells scattered around. Additionally, we see great amounts of mono-vacancies somewhat uniformly distributed and in some cases we can see vacancy clusters forming stacking-fault tetrahedra (SFT). In all cases, we see dislocation structures consisting of Shockley-type dislocations. Frank loops and dislocation segments were also observed during the overlapping cascades. However, these were mostly short-lived and were seen most often in AlCrCuFeNi$_3$.

\begin{figure}[ht!]
    \centering
    \includegraphics[width=0.9\linewidth]{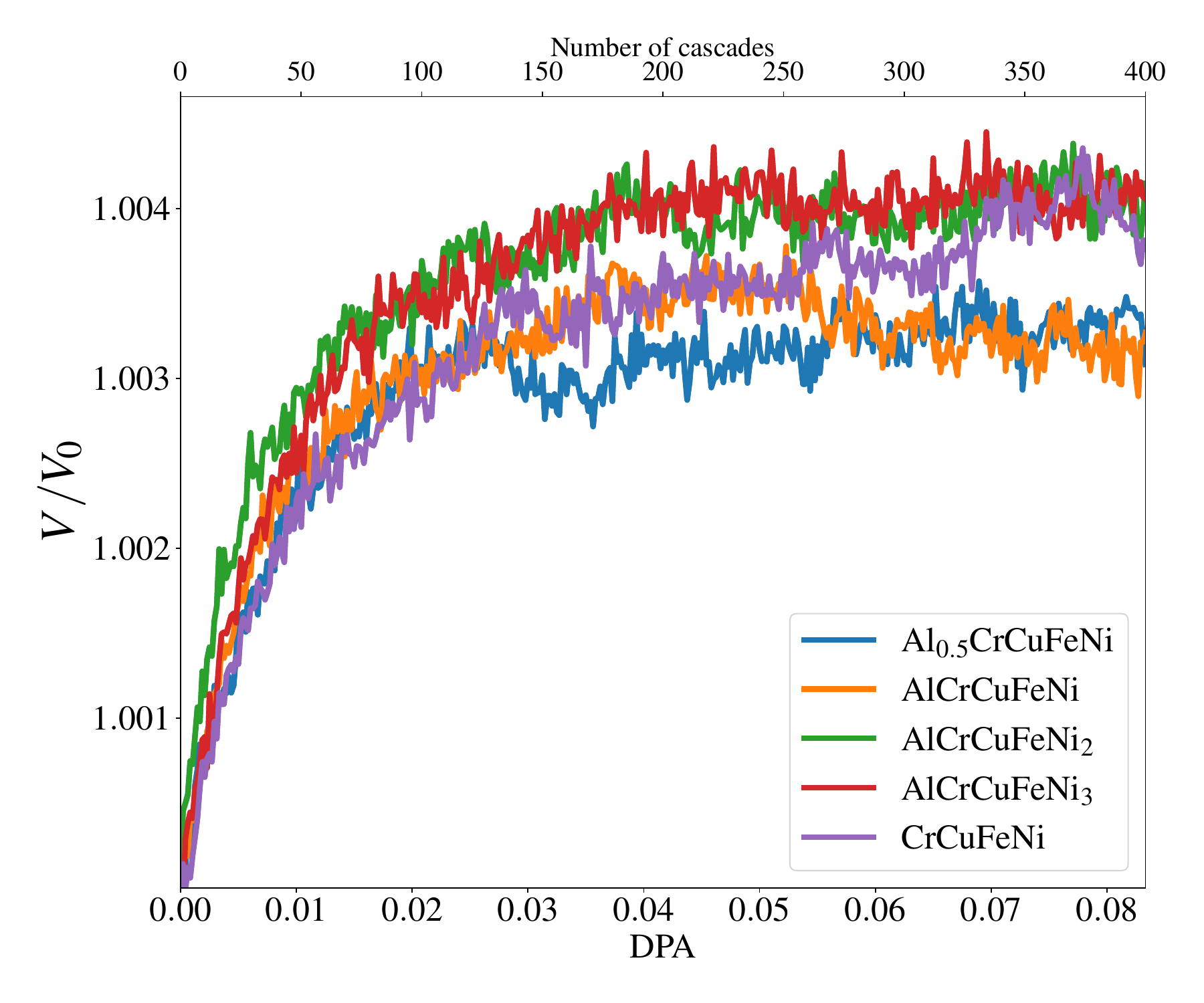}
    \caption{Volume change as a function of dose for the different compositions. }
    \label{fig:volume}
\end{figure}

Fig.~\ref{fig:volume} shows the volume change as a function of the dose. Interestingly, there is no clear correlation with volume change and the defect concentration, as one could intuitively assume. We can see that the composition with the highest defect concentration is one of the lowest in volume change and conversely the composition with the lowest defect concentration is one of the highest. Similar behavior has been reported in tungsten-based alloys, where factors such as cluster size and morphology were discussed to play a key role in volumetric swelling in addition to defect concentration~\cite{wei2024revealing}. 

\begin{figure}
    \centering
    \includegraphics[trim={0 0.5cm 0 0.5cm},clip,width=0.95\linewidth]{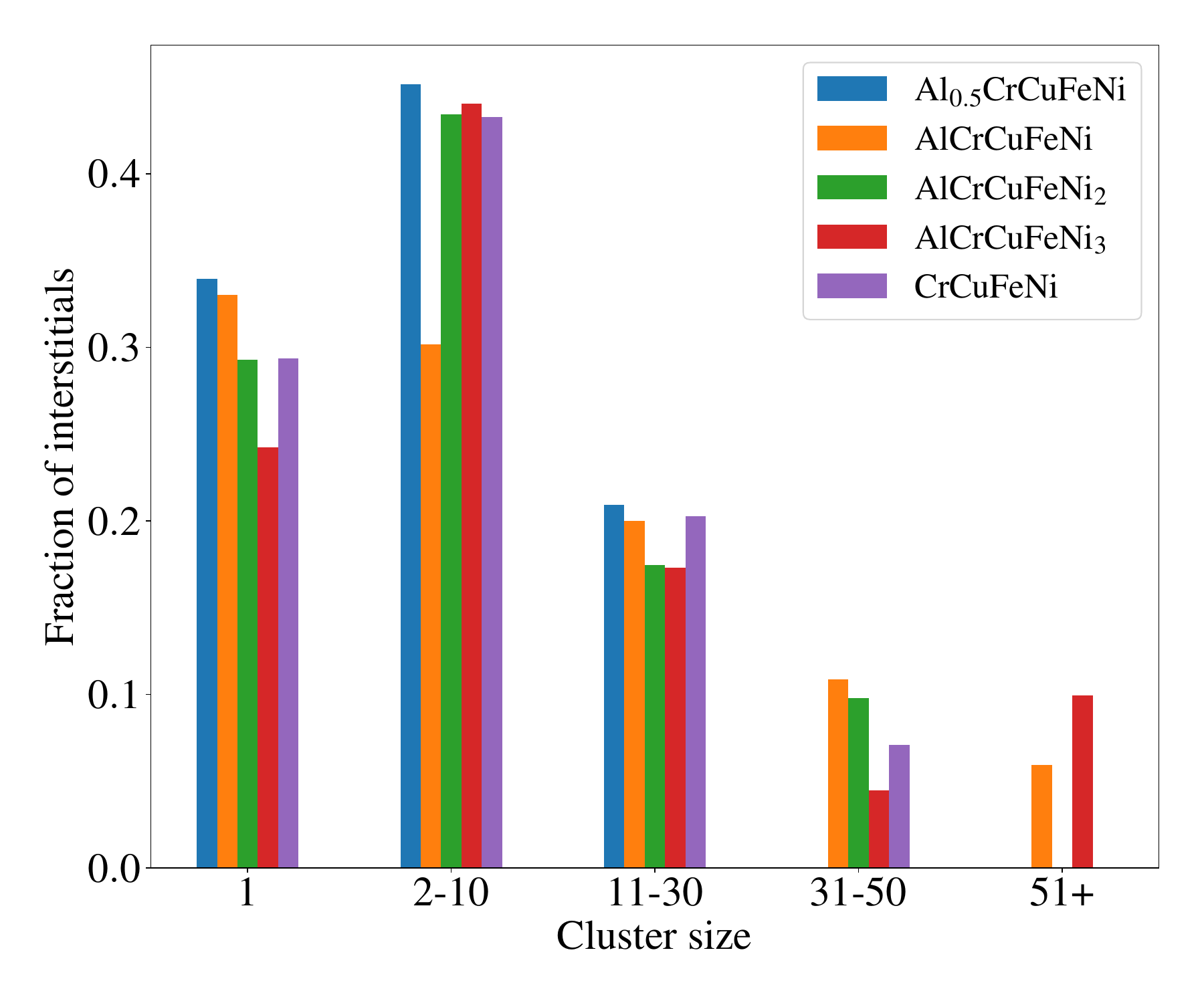}
    \includegraphics[trim={0 0.5cm 0 0.5cm},clip,width=0.95\linewidth]{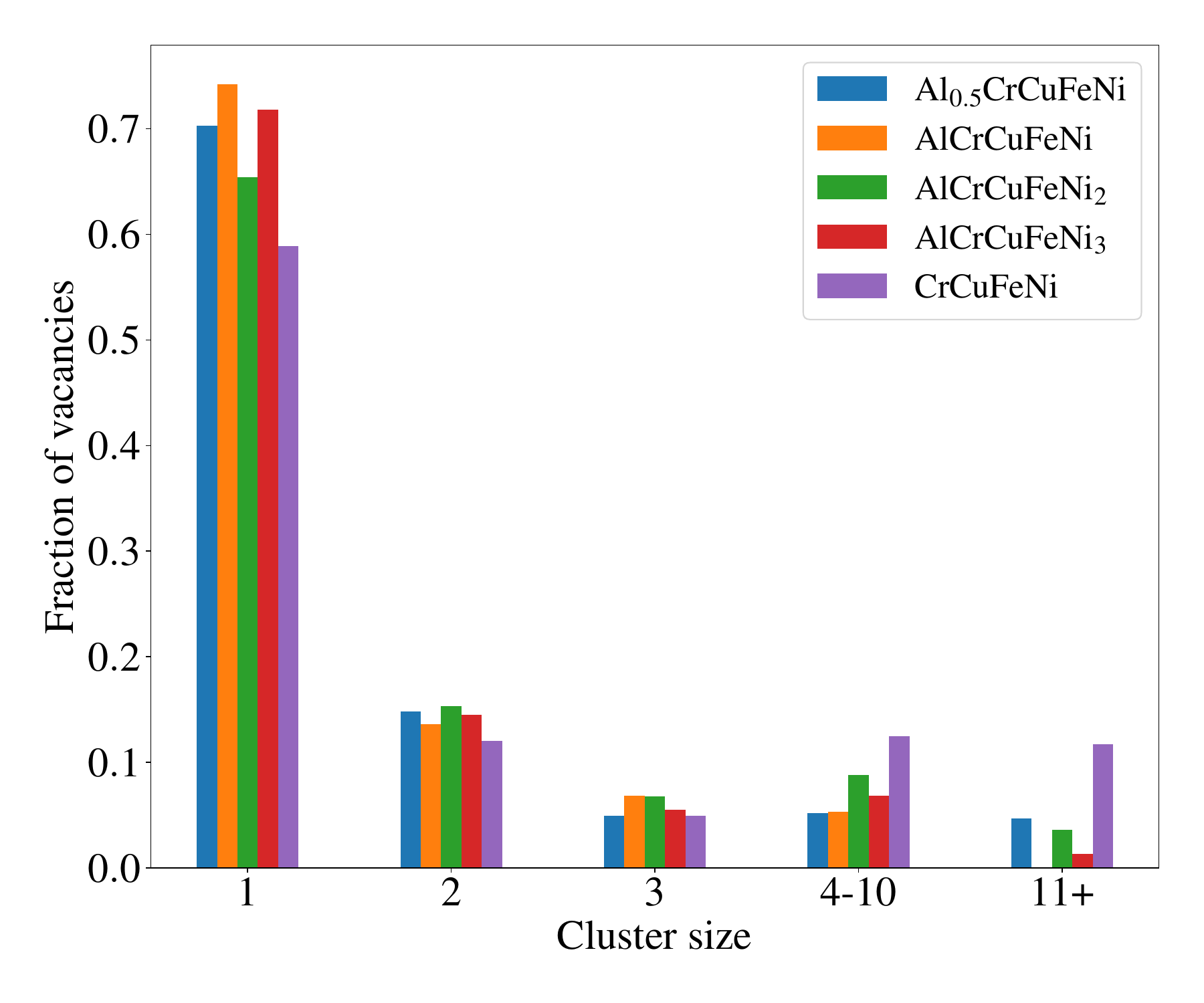}
    \caption{Total number of interstitials and vacancies in clusters of different sizes. } 
    \label{fig:CA}
\end{figure}

Fig.~\ref{fig:CA} shows the results for both interstitial and vacancy clusters. Comparing the results from the cluster analysis to the defect concentrations, we can see that the compositions with the highest defect concentration also formed the largest interstitial clusters. Looking at the vacancies, we see that the majority of the vacancies are mono-vacancies and the clusters are overall smaller than for interstitials. The compositions that exhibited the highest volumetric swelling also had proportionately more larger vacancy clusters, which might be a key contributing factor and is consistent with what has been reported in tungsten-based alloys~\cite{wei2024revealing}.

\begin{figure}[ht!]
    \centering
    \includegraphics[width=0.95\linewidth]{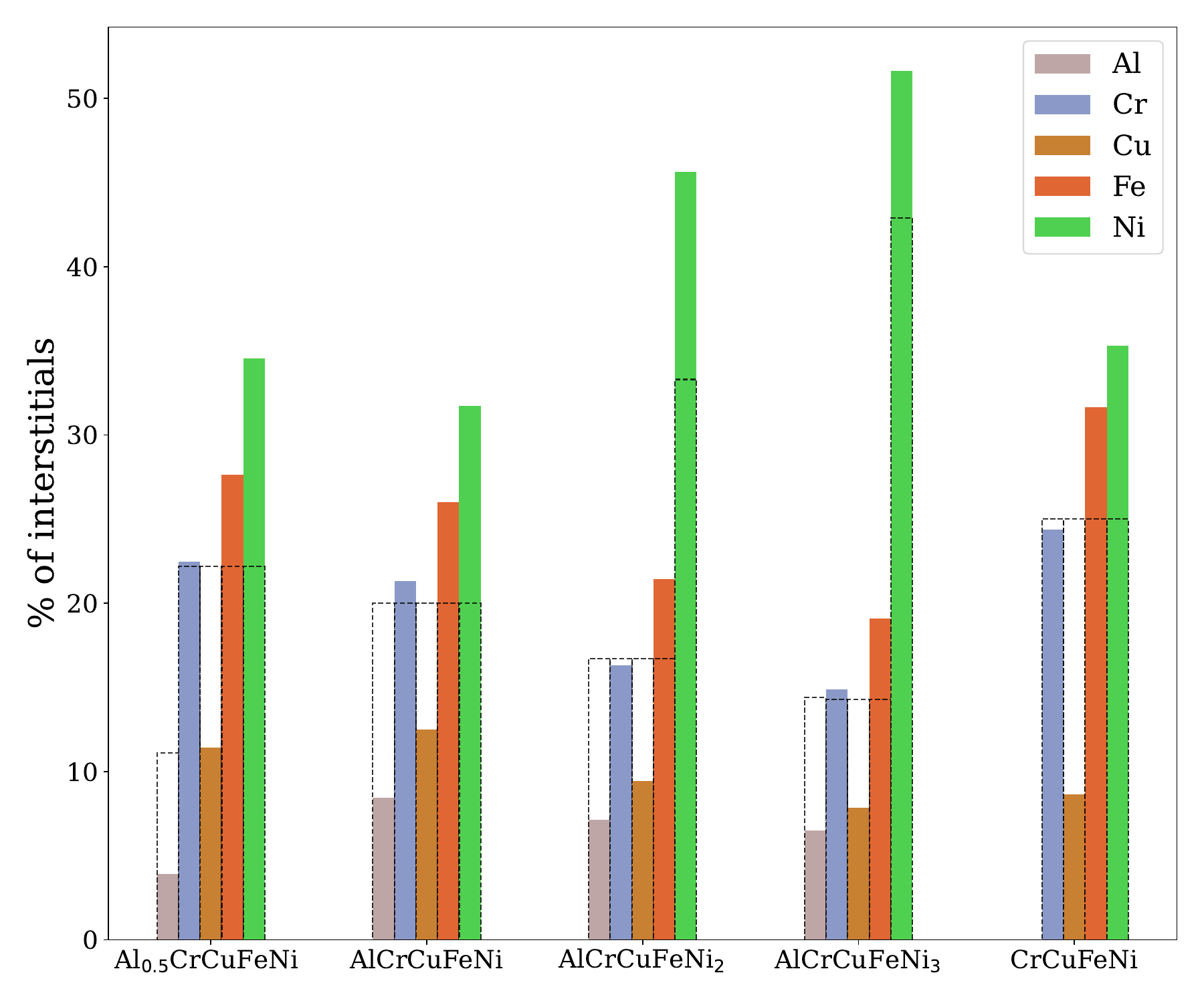}
    \caption{Chemical types of interstitial after 400 cascades. Dashed outline is the estimation assuming equal probability between chemical species.}
    \label{fig:Chem-type}
\end{figure}

\begin{figure} [ht!]
    \centering
    \includegraphics[width=0.95\linewidth]{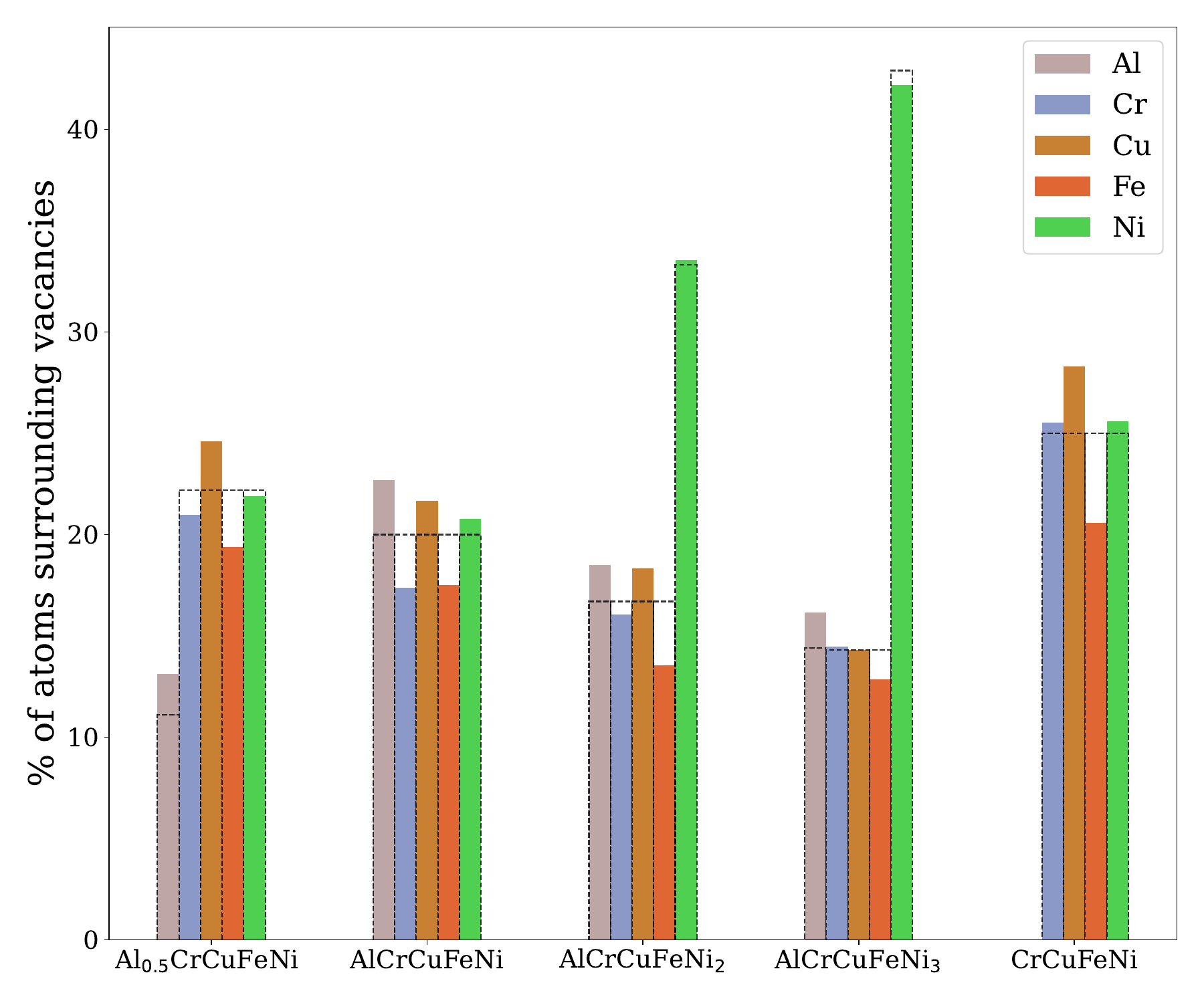}
    \caption{Chemical types of atoms around vacancies after 400 cascades. Dashed outline is the estimation assuming equal probability between chemical species.}
    \label{fig:vac-type}
\end{figure}

Fig.~\ref{fig:Chem-type} shows the chemical species of the interstitial atoms of the different HEA compositions, where the expected percentage of interstitials assuming equal probability between chemical types is also imposed. Ni and Fe are clearly overrepresented in the interstitials, while Al and Cu are clearly underrepresented, which is the case for all compositions. In Fig.~\ref{fig:vac-type} we have done similar analysis but with the atoms surrounding vacancies. In this case, we have counted the chemical types of the atoms that are within a certain radius ($r_{\mathrm{2.5nn}}$) of each vacancy, making sure not to count atoms multiple times. In contrast to interstitials, Al and Cu are overrepresented in the atoms surrounding the vacancies. However, this effect is not that significant. Additionally, there is depletion of Fe in the atoms surrounding vacancies.

\begin{figure} [ht!]
    \centering
    \includegraphics[width=0.9\linewidth]{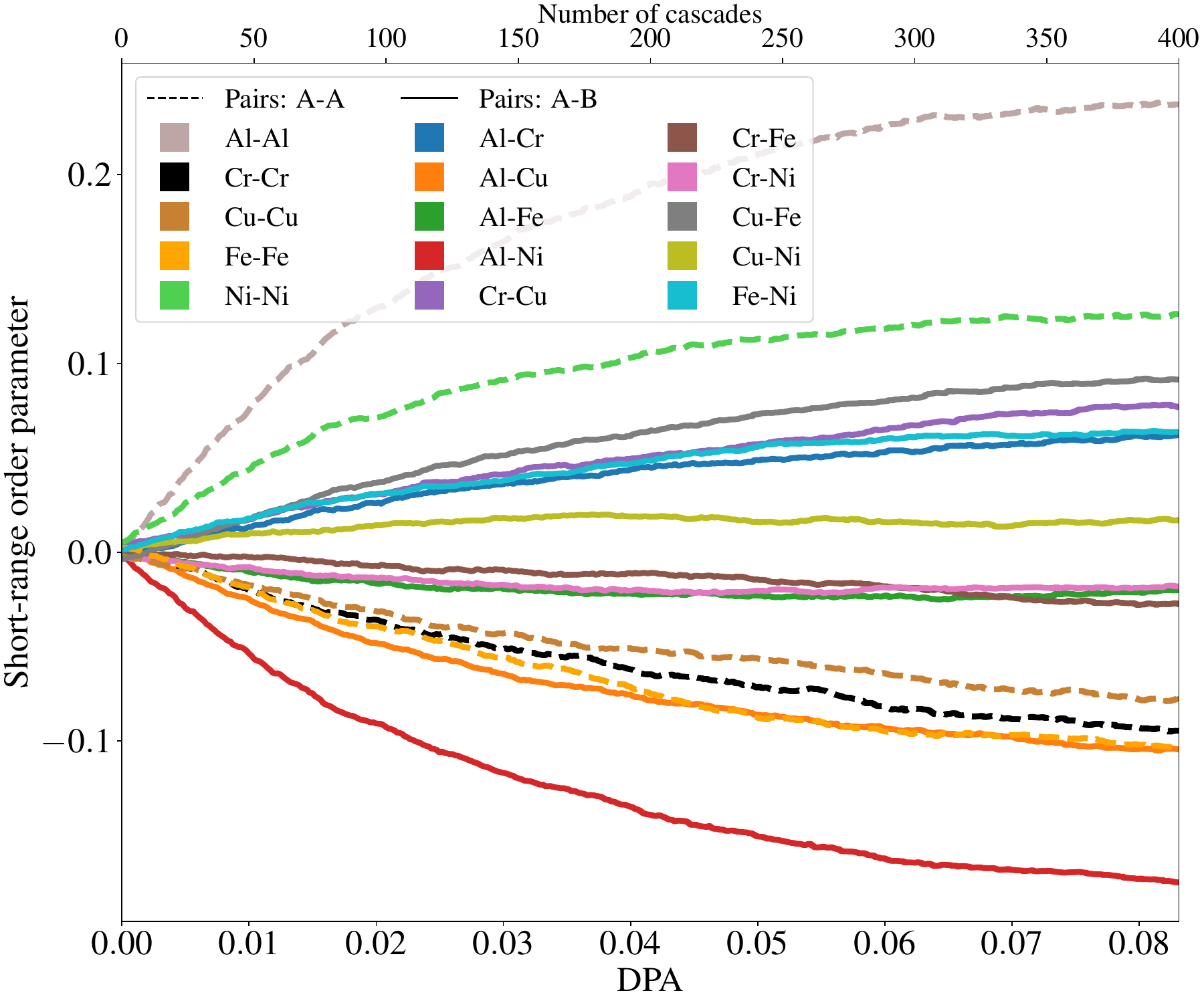}
    \caption{Short-range order parameters for all element pairs in AlCrCuFeNi during the overlapping cascades. The dashed lines represent all pair with the same chemical type and the others are between pairs of different chemical types.}
    \label{fig:sro}
\end{figure}

Fig.~\ref{fig:sro} shows the Warren-Cowley short-range order (SRO) parameters~\cite{cowley1950approximate} (first shell) calculated for all atoms pairs in the equiatomic composition. It is immediately evident that there is preferential ordering that is induced as a consequence of the collision cascades. The negative values mean a tendency for short-range ordering, while positive values indicate the opposite tendency. Al-Al and Ni-Ni pairs show the strongest tendency to separate, while Al-Ni shows strongest tendency to order. This is logical, as it means that the Al-Al and Ni-Ni separate to form Al-Ni ordering. The Cu-Cu, Fe-Fe and Cr-Cr elemental pairs all show a degree of short-range ordering, with Cu also showing significant ordering with Al. Furthermore, we see that the SRO parameters show some degree of saturation although they are still changing well after defect concentration has stabilized (see Fig.~\ref{fig:defect-concetration}, around 0.04--0.05 dpa). There are differences in the values of the SRO parameters and ordering of the pairs between the different compositions. However, the overall tendencies are similar. In all cases (that include Al), Al-Al pairs show strongest tendency to separate, while Al-Ni shows strongest tendency for ordering. In the CrCuFeNi composition, Cu-Cu and Fe-Fe pairs show the strongest tendency for ordering. The SRO parameters of the other compositions can be found in the supplementary materials (Fig. S4). 

\subsection{Annealing}

\begin{figure}[ht!]
    \centering
    \includegraphics[width=0.95\linewidth]{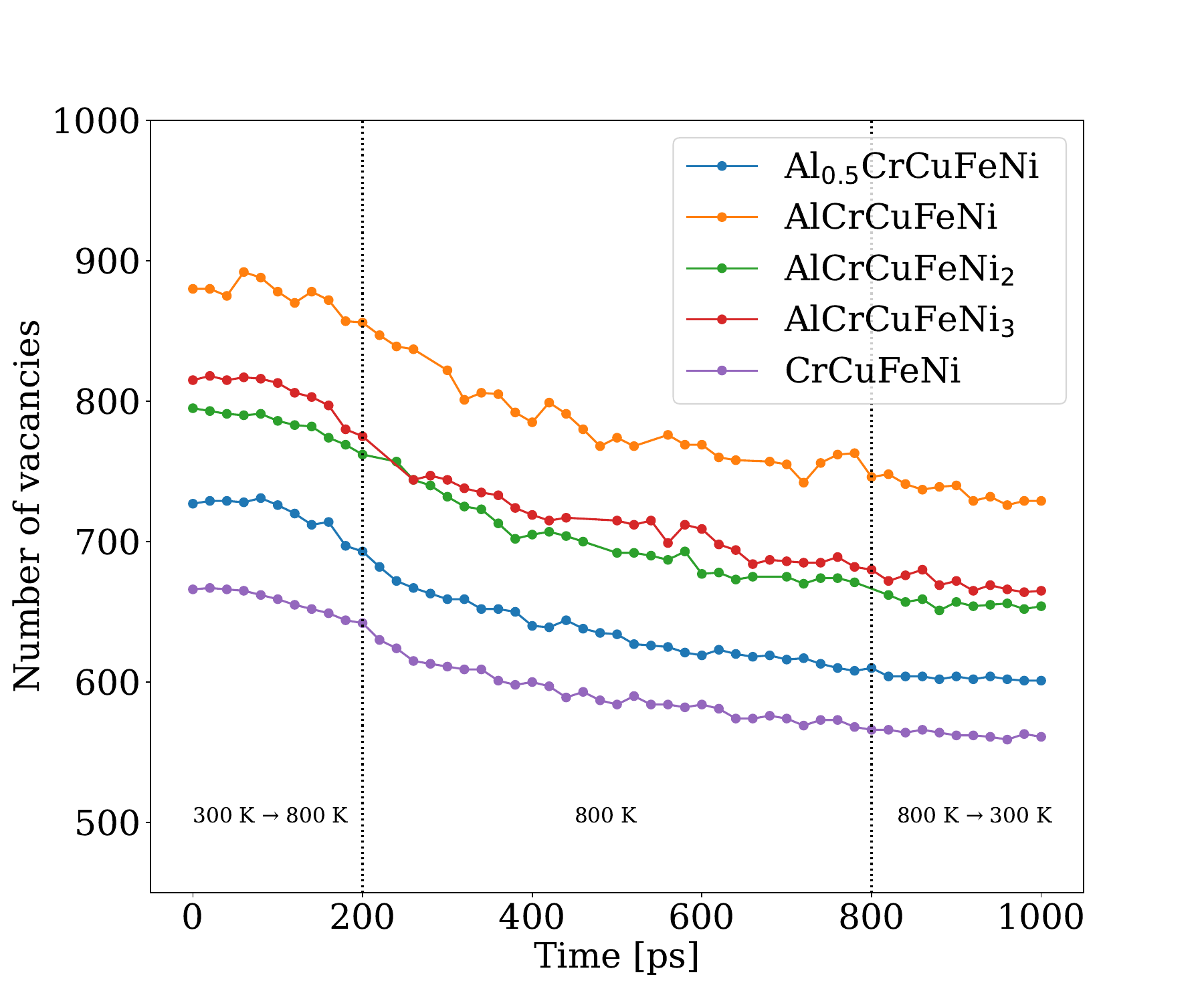}
    \caption{Number of vacancies during annealing at 800 K for different HEA compositions.}
    \label{fig:anneling}
\end{figure}

To accelerate thermal recombination that cannot be accessed directly within MD time scales, MD annealing runs at elevated temperatures are frequently applied. For tungsten-based alloys, the defect structures in simulation cells of different compositions were shown to exhibit significant differences after annealing, which was explained by the different diffusion rates of interstitial and vacancy-type defects~\cite{wei2024revealing}. To investigate whether this effect is also present in the FCC HEAs which are considered in this work, the final irradiated structures were annealed for 1 ns at 800 K. The temperature was chosen to be high enough for defect recombination to take place at an accelerated rate, while still being well below the melting temperatures of the materials. The annealing was performed in the $NPT$ ensemble and consisted of three stages. First, the temperature of the system was raised from 300 K to 800 K during the first 200 ps. Secondly, the system was held at 800 K for the next 600 ps, and finally the system was cooled back to the initial 300 K during the final 200 ps of the simulation. During this process, the vacancy concentration of the systems were tracked. Fig.~\ref{fig:anneling} shows the vacancy concentration as a function of time during annealing.  During the simulation, roughly 17 \% of the vacancies recombine and anneal out. However, significant differences between the different compositions were not observed, at least at this temperature and time scale. Additionally, the final structures after annealing, while not identical, did not show significant differences between the compositions. The final structures can be found in the supplementary materials (Fig. S5).

\section{Discussion}

Our general-purpose interatomic potential developed for AlCrCuFeNi alloys enables computationally efficient and accurate computational investigations of various alloys. While the resulting tabGAP model does not explicitly include magnetic degrees of freedom, the use of magnetic DFT training data is key to enable simulations of multiphase alloys and phase transitions. This is evident from the Monte Carlo swapping MD simulations, where a dynamic transition from FCC to BCC was observed for the equiatomic composition. Magnetism has been studied  experimentally recently in the  AlCrCuFeNi alloy, where it was noted the alloys exhibit robust hard ferromagnetism~\cite{yazdani2024tribological}. Furthermore, L. Kush et al. have estimated the Curie temperature of the alloy between 341.7 to 404.9 K depending on the composition~\cite{kush2022effect,kush2020structural}. It is also stated that below the Curie temperature the alloy has a very stable ferromagnetic order. This means that the assumption of the initial ferromagnetic order in the DFT training data is acceptable. However, at higher temperatures or due to compositional variations and ordering, the alloy may exhibit more complex magnetic behavior, limiting the transferability of the current model. In such cases a more accurate model that takes into account the explicit magnetic degree of freedom will be required. Even more strikingly, the MCMD simulations computationally support the VEC stability rule (see Table~\ref{tab:compositions} for VECs), where for alloy compositions with VEC $\geq 8$ a single FCC phase is expected and for $6.87 \leq$ VEC $ <$ 8 FCC and BCC coexistence is expected~\cite{guo2011effect}. This rule has also found support in experiments~\cite{guo2013anomalous,suliz2022synthesizing,oh2022controlled}. However, the VEC rule might not be well suited for prediction of phases in non-equilibrium rapidly solidified HEAs~\cite{luo2021microstructural}, and BCC phases have been observed with AlCrCuFeNi alloys above the VEC criterion~\cite{kafali2023wear}. Furthermore, the addition of large amounts of Al causes significant lattice distortions (due to it having a large atom radius, see Table~\ref{tab:compositions} for $\delta\%$), which makes the close-packed FCC less stable and thus the BCC lattice can accommodate the larger Al atoms more easily in it~\cite{tang2013aluminum}. 

The overlapping cascade simulations provide insight into the alloys' response to irradiation and also reveal a strong Al dependency with regards to the accumulated defect concentration. Generally, we found that alloys with proportionally more Al accumulates higher defect concentrations. The large lattice distortions induced by Al might play a part in this. However, this is not a one-to-one correlation, as the defect concentrations do not directly follow the atomic size differences $\delta\%$ of the compositions. Another possible reason is the mass differences between the elements, where Al has a significantly smaller mass than the rest. To investigate the mass effect, we performed an unphysical test simulation where we substituted the Al mass with the mass of Ni in the equiatomic composition, while keeping everything else the same. In this unphysical case, the defect concentration increased slightly (enhancing the Al effect), so the difference cannot be attributed to the mass difference, and must be due to chemical effects and lattice distortion. The defect concentration in this unphysical case can be found in the supplementary materials (Fig. S6). During continued irradiation, we observed clear short-range ordering as a consequence of the atomic mixing from the overlapping cascades. The SRO parameters approach saturation after continued irradiation. This is in line with a previous computational study in CuNiCoFe, where a steady-state condition was observed for short-range ordering~\cite{koch2017local}. There it was observed that starting from initially short-range-ordered or random states resulted in similar SRO parameters after continued irradiation.

Our results reaffirms, that machine-learned interatomic potentials have truly expanded the capabilities of atomistic simulations both in terms of scale and accuracy. Previously, fitting an analytical potential for complex alloys with many different chemical species was cumbersome and as the multi-element fitting often relied on rather simple mixing rules, it was not clear whether the potential could be expected to provide realistic results or not. The ML approach (while not devoid of weaknesses and challenges) provides a clear path to building interaction models for complex alloys. Furthermore, the DFT databases used in the training can be further amended and extended, when for example new material properties or additional elements need to be considered. Extensive validation is still needed in the case of machine-learned interatomic potentials, which can be challenging especially for complex alloys. Unfortunately, the previous MD studies in AlCrCuFeNi alloys with mixed EAM and Morse potentials~\cite{li2016atomic, li2016mechanical,wang2017investigation,zeng2019thermal,niu2021molecular,doan2021effects,nguyen2023plastic,nguyen2023cyclic,nguyen2024machining,doan2025mechanical} report a very limited amount of validation, inhibiting a retroactive assessment of the accuracy of the previous results.

\section{Conclusions}

In this work, we have developed a machine-learned interatomic potential (tabGAP) for the AlCrCuFeNi high-entropy alloy system, taking magnetism implicitly into account by training to spin-polarized DFT calculations. This is to the authors' knowledge the first general-purpose interatomic potential for the AlCrCuFeNi system. The potential is suitable for large-scale simulations and is developed with additional considerations for radiation damage simulations. Using the potential, Monte Carlo swapping simulations were performed in order to investigate phase stability and preferential ordering and segregation in the HEAs. We observed a strong tendency for Fe-Cr and Cu segregation, which is in line with experimental observations. Additionally, we observed a phase transition from FCC to BCC initiated by Fe-Cr precipitation in the equiatomic compositions, which is in line with the proposed valence electron concentration stability rule. Furthermore, we used the potential to simulate massively overlapping collision cascades in different compositions of the HEA, where we found that the defect concentration showed clear Al concentration dependency. Additionally, we saw a clear emergence of short-range ordering during continuous irradiation.

\section*{Acknowledgments}

This work has received funding from the Academy of Finland through the HEADFORE project (grant number 333225).
The authors wish to thank the Finnish Computing Competence Infrastructure (FCCI) and CSC -- IT Center for Science for supporting this project with computational and data storage resources.
This work has been partially carried out within the framework of the EUROfusion Consortium, funded by the European Union via the Euratom Research and Training Programme (Grant Agreement No 101052200 — EUROfusion). Views and opinions expressed are however those of the author(s) only and do not necessarily reflect those of the European Union or the European Commission. Neither the European Union nor the European Commission can be held responsible for them.

\bibliography{mybib}

\end{document}